\documentclass[
superscriptaddress,
 amsmath,amssymb,
 aps,
]{revtex4-2}

\usepackage{graphicx}% Include figure files
\usepackage{dcolumn}% Align table columns on decimal point
\usepackage{bm}% bold math
\usepackage{appendix}
\usepackage{subfigure}
\usepackage[mathlines]{lineno}% Enable numbering of text and display math
\newcommand {\pt}{p_{\rm T}}
\newcommand {\s}{\sqrt{s}}
\begin{document}

\title{PYTHIA 8 underlying event tune For RHIC energies}
\author{Manny Rosales Aguilar}
\affiliation{University of Kentucky, Lexington, Kentucky 40506, USA}
\author{Zilong Chang}
\affiliation{Brookhaven National Laboratory, Upton, New York 11973, USA}
\author{Raghav Kunnawalkam Elayavalli}
\affiliation{Yale University, New Haven, CT 06517, USA}
\affiliation{Brookhaven National Laboratory, Upton, New York 11973, USA}
\author{Renee Fatemi}
\affiliation{University of Kentucky, Lexington, Kentucky 40506, USA}
\author{Yang He}
\affiliation{Shandong University, Qingdao, Shandong 266237, China}
\author{Yuanjing Ji}
\affiliation{Lawrence Berkeley National Laboratory, Berkeley, CA 94720, USA}
\author{Dmitry Kalinkin}
\affiliation{Indiana University, Bloomington, Indiana 47408, USA}
\author{Matthew Kelsey}
\email{mkelsey@wayne.edu}
\affiliation{Wayne State University, Detroit, MI 48202, USA}
\author{Isaac Mooney}
\affiliation{Wayne State University, Detroit, MI 48202, USA}
\author{Veronica Verkest}
\affiliation{Wayne State University, Detroit, MI 48202, USA}
\date{\today}

\begin{abstract}
We report an underlying event tune for the PYTHIA 8 Monte Carlo event generator that is applicable for hadron collisions primarily at $\s$ ranges available at the Relativistic Heavy-Ion Collider (RHIC). We compare our new PYTHIA 8 tuned predictions to mid-rapidity inclusive $\pi^{\pm}$ spectra, jet sub-structure, Drell-Yan production, and underlying event measurements from RHIC and the Tevatron, as well as underlying event data from the Large Hadron Collider. With respect to the default PYTHIA 8 Monash Tune, the new `Detroit' tune shows significant improvements in the description of the experimental data. Additionally, we explore the validity of PYTHIA 8 predictions for forward rapidity $\pi$ in $\s$ = 200 GeV collisions, where neither tune is able to sufficiently describe the data. We advocate for the new tune to be used for PYTHIA 8 studies at current and future RHIC experiments, and discuss future tuning exercises at lower center-of-mass energies, where forward/backward kinematics are essential at the upcoming Electron-Ion collider. 
\end{abstract}

\maketitle

\section{Introduction\label{sec:intro}}

Monte Carlo (MC) event generators that simulate relativistic lepton-lepton, lepton-hadron, and hadron-hadron collisions are an essential part of high-energy particle and nuclear physics. From a theoretical perspective, MC models are able to test our fundamental understanding of the Standard Model, in particular Quantum Chromodynamics (QCD), and offer a prescription for the initial and final states of the collision system. From the experimental point of view, MCs are an integral piece in the simulation chain that aims to reproduce realistic spectra that are ultimately used to extract detector acceptance and resolution corrections, and study systematic effects in experimental data. There are several event generators that are currently available, for example PYTHIA~\cite{Sjostrand:2006za,Sjostrand:2007gs,Sjostrand:2014zea} and Herwig/Herwig++~\cite{Bellm:2015jjp,Bahr:2008pv}, that have been widely used to simulate collisions at the Large Hadron Collider (LHC) and the Relativistic Heavy-Ion Collider (RHIC). The simulation routines include various QCD physics processes that factorize a single collision into two regimes consisting of the perturbative hard scattering and evolution via a parton shower, and various non-perturbative components such as hadronization, underlying event and multi-parton interactions. 

Physics processes implemented in event generators include multiple parameters which are turned to experimental measurements, often from $e^{+}+e^{-}$ collisions. The PYTHIA event generator studied in this publication has been very successful in describing data at the LHC and has been the subject of many tuning exercises over the past decades~\cite{PhysRevD.82.074018,Skands:2014pea,Sirunyan2020,Gunnellini2018,ATLAS:2012uec,Buckley2009,CMS:2015wcf,TheATLAScollaboration:2014rfk}. While the global tuning of PYTHIA 6 and 8 in Refs.~\cite{PhysRevD.82.074018,Skands:2014pea} are in good agreement with data at LHC energies, there are significant discrepancies in describing data from collisions at lower center-of-mass energies~\cite{PhysRevD.103.L091103,PhysRevD.101.052004}. These disagreements can mainly be understood as a consequence of incorrect modeling of the soft QCD underlying event (UE) stemming from the center-of-mass energy extrapolation that is used. A PYTHIA 6 ``STAR" tune was produced in Ref.~\cite{STAR:2019yqm} which updated the value of the parameter controlling the energy extrapolation of the low transverse momentum ($\pt$) cross section, and was able to adequately describe the $\pi^{\pm,0}$ $\pt$ and jet $\pt$ spectrum in proton-proton ($p$+$p$) collisions at $\sqrt{s} = 200$ GeV from STAR and PHENIX as demonstrated in Refs.~\cite{PhysRevD.103.L091103,PhysRevC.102.054913}. For PYTHIA 8, there has been several tuning exercises with LHC data at $\s=$ 7 and 13 TeV collisions, and some lower collision energy data either from the LHC or the Tevatron~\cite{CMS:2015wcf,Gunnellini2018,ATLAS:2012uec}. The LHC-focused tune produced in~\cite{ATLAS:2012uec} utilized LHC data from $\s$ = 900 GeV and 7 TeV $p$+$p$ collisions, and it was argued therein that including Tevatron data within the tuning procedure leads to inadequate tune performance at LHC energies. In contrast, measurements of the UE multiplicity from CDF~\cite{PhysRevD.92.092009} at $\s$ = 900 and 1960 GeV and LHC data at the top energies $\s$ = 7 and 13 TeV were used in this~\cite{CMS:2015wcf} tuning exercise to control the impact of energy extrapolation. However, the aforementioned tune was not able to describe the CDF UE data at $\s$ = 300 GeV. The energy dependence of low $\pt$ regularization was also studied in~\cite{Gunnellini2018} using the same LHC data and CDF data at $\s$ = 300, 900, and 1960 GeV, and concluded that the inclusion of a new term in the extrapolation produced improved agreement at lower energies. All these observations motivate the need for a dedicated PYTHIA 8 tune, in addition to the existing PYTHIA 6 STAR tune, for the nominal RHIC energy of $\s$ = 200 GeV.    

The organization of the following sections is as follows: section~\ref{sec:method} describes the general tuning procedure via the Professor toolkit~\cite{Buckley2009}; section~\ref{sec:data} is dedicated to describing the data and RIVET~\cite{10.21468/SciPostPhys.8.2.026} implementation; section~\ref{sec:newtune} presents our tuned results; section~\ref{sec:comp} compares the new PYTHIA 8 predictions to selected data distributions; in Section~\ref{sec:forward} we compare the default and tuned distributions at forward rapidity; and section~\ref{sec:conclusion} summarizes our results. The Appendix~\ref{app:a} presents the comparison of our new tune to all the experimental data.

\section{Tuning procedure\label{sec:method}}

We utilize a parameterization-based approach for the tuning procedure provided by the Professor (v2.3.3)~\cite{Buckley2009} toolkit. In the Professor tuning methodology, PYTHIA parameters of interest are sampled $n$ times across a provided range and for each sampling a MC generation is produced and the resulting prediction is compared to data. In each bin of data, the result of the random samplings are parameterized by a third order polynomial. In the case of $N$ PYTHIA tuning parameters, the corresponding polynomials are $N$-dimensional. The coefficients of the polynomials are computed numerically within the Professor code. A $\chi^{2}$ fit of the polynomial parameterizations to the data is performed using Minuit~\cite{iminuit,James:1975dr} to determine the best values of each tuning parameter.    

The starting point of our tuning exercise is the default Monash tune~\cite{Skands:2014pea} and PYTHIA 8.303. The NNPDF2.3 parton distribution functions (PDF)~\cite{Ball:2012cx} used in the Monash tune have since been updated with improved data and methods, and therefore we utilize the recent NNPDF3.1 leading-order PDF set with $\alpha_{s}(m_{Z})$ = 0.130~\cite{Ball2017}. The settings that are varied in our tune are $p_{T,0}^{Ref}$ and its energy-dependence scaling parameter, the proton matter distribution parameters, and color reconnection range. The $p_{T,0}^{Ref}$ parameters regularizes the low $p_{T}$ cross section divergence, and is a key parameter in all PYTHIA tunes. The energy-scaling of $p_{T,0}$ follows a power-law function, and is controlled by the ecmPow parameter. We change the reference energy that corresponds to the $p_{T,0}^{Ref}$ parameter to be 200 GeV. We do so for two reasons: 1) using a different PDF set requires a complete re-tune of $p_{T,0}^{Ref}$; and 2) to control the power-law extrapolation as much as possible at low center-of-mass energies due to the rapidly varying functional form in this region. For the proton shape function, we change it to the double Gaussian matter profile (MultipartonInteractions:bProfile=2) and vary it's two respective parameters, coreRadius and coreFraction. Table~\ref{tab:params} tabulates all the tuning parameters and their respective ranges. From the previous PYTHIA 6 "STAR" tune study in~\cite{STAR:2019yqm}, the values for $p_{T,0}^{Ref}$ and ecmPow are expected to be smaller than the default values, and therefore their ranges are chosen to cover all values lower than the default with some overlap. The other tuning parameter ranges cover all possible values. A complete description of the above tuning parameters are given in~\cite{Sjostrand:2006za,Sjostrand:2007gs,Sjostrand:2014zea}  

\begin{table}[!htbp]
	\caption{PYTHIA 8 settings and tuning parameters. \label{tab:params}}
	\centering
	\begin{tabular}{c | c | c}
 	\hline \hline
    Setting & Default & New  \\ \hline	
    PDF:pSet & 13 & 17 \\
    MultipartonInteractions:ecmRef & 7 TeV & 200 GeV \\
    MultipartonInteractions:bprofile & 3 & 2 \\   
    \hline
    Tuning Parameter & Default & Range  \\ \hline
    MultipartonInteractions:pT0Ref & 2.28 GeV &  0.5-2.5 GeV     \\    
    MultipartonInteractions:ecmPow & 0.215 & 0.0-0.25      \\   
    MultipartonInteractions:coreRadius & 0.4 & 0.1-1.0      \\ 
     MultipartonInteractions:coreFraction & 0.5 & 0.0-1.0      \\   
    ColourReconnection:range       & 1.8 & 1.0-9.0      \\   
	\hline \hline
		\end{tabular}
	
\end{table}

We sample 70 values of the tuning parameters in Table~\ref{tab:params} within the specified ranges using the Professor package to be used as anchor points in the generator response polynomial parameterization. We have checked that using a smaller fraction (but still sufficient for 5 tuning parameters) of the sampled anchor points produces compatible results with respect to the full sample. We generate 10 million events for each simulation run to ensure the MC statistics are sufficiently less than that of the data in the region of interest. The tuned results are determined by minimizing the weighted $\chi^{2}$,
\begin{equation}
\chi^{2} = \sum_{i} w_{i}(F[i] - d_{i})C^{-1}(F[i] - d_{i}),
\end{equation}
where the index $i$ runs over all data points $d_{i}$, $w_{i}$ is a weight for each data point, $F[i]$ is the parameterized PYTHIA prediction for the $i$-th data point, and $C^{-1}$ is the inverse covariance matrix which contains only experimental uncertainties (assumed no bin-by-bin correlations). For the final fit we do not assign any weights beyond unity to any of the data. 

\section{Input data and MC generation\label{sec:data}}

Mid-rapidity data from STAR, PHENIX, and CDF are utilized in this tuning exercise. The measurements can be grouped in three main categories; identified particle spectra, event multiplicities, and jet substructure as described in Table~\ref{tab:data}. Collectively, they broadly sample the available phase space and test the model's ability to describe the non-perturbative and perturbative parts of a $p$+$p$ collision. The spectra measurements focus on $\pi^{\pm}$ yields at STAR~\cite{ADAMS2006161} and Drell-Yan di-muon pair production from PHENIX~\cite{PhysRevD.99.072003}. We include measurements of the UE event multiplicity from STAR~\cite{PhysRevD.101.052004} and CDF~\cite{PhysRevD.92.092009} at various regions such as towards, away, and transverse with respect to the trigger object, be it a reconstructed jet or charged hadron, respectively. 
Lastly, two sets of jet sub-structure measurements from STAR~\cite{ADAM2020135846,starcollaboration2021invariant} on the SoftDrop splitting observables at the first split and the invariant and groomed jet mass, all measured differentially as a function of the jet $\pt$ and jet resolution parameter $R$, are included in the tuning exercise. The substantial impact on jet substructure observables from the variation of the MPI and UE parameters is an indication of the multi-faceted inner workings of the PYTHIA event generator. The last column of Table~\ref{tab:data} mentions the respective figure in both the paper draft and in the appendix where the comparisons with the data are shown. 

\begin{table*}[!htbp]
	\caption{Mid-rapidity data used in the tuning procedure. \label{tab:data}}

	\centering
	\begin{tabular}{c | c | c | c | c}
		\hline \hline
    Experiment & $\sqrt{s}$ (GeV) & Observable & Reference & Figure \\ \hline	
    STAR & 200 & $\pi^{\pm}$ cross sections vs. $p_{T}$ & \cite{ADAMS2006161} & ~\ref{fig:datacomp},~\ref{fig:pid} \\
    PHENIX & 200 & Di-muon pairs from Drell-Yan vs. di-muon $p_{T}$  & \cite{PhysRevD.99.072003} &~\ref{fig:dy} \\
    \hline
    STAR & 200 & Average charged particle multiplicities and $p_{T}$ vs. leading jet $p_{T}$ & \cite{PhysRevD.101.052004} &~\ref{fig:datacomp},~\ref{fig:ue1},~\ref{fig:ue2} \\
     &  & in the forward, transverse, and away regions & &   \\
    CDF & 300,~900,~1960 & Charge particle density and $\sum p_{T}$ vs. leading hadron $p_{T}$ in & \cite{PhysRevD.92.092009} &~\ref{fig:cdfue1},~\ref{fig:cdfue2},~\ref{fig:cdfue3} \\
         &  & transverse region & &  \\
    \hline
    STAR & 200 & SoftDrop groomed jet sub-structure ($z_{\rm{g}}$ and $R_{\rm{g}}$) & \cite{ADAM2020135846} &~\ref{fig:jet}\\    
    STAR & 200 & Inclusive and groomed jet mass & \cite{starcollaboration2021invariant} &~\ref{fig:datacomp},~\ref{fig:jet}\\
   	\hline \hline         
		\end{tabular}
\end{table*}

We prepared RIVET analyses for each of the measurements mentioned above (when analyses were not publicly available) and they are all made available here \href{https://github.com/star-bnl/star-pythia8-tune}{github.com/star-bnl/star-pythia8-tune}. 

The various MC runs are all analyzed by the RIVET analyses and the resulting output yoda files are processed by the Professor toolkit to determine the minimum $\chi^{2}$. We exclude the $\pi^{\pm}$ cross section below $p_{T}< 1$ GeV/$c$ to avoid potential feed-down effects which we studied by turning weak decays on or off. We also remove data from the fit in which the envelope of the PYTHIA predictions are significantly varied compared to the data systematic uncertainties. Similarly, data points where our MC statistical uncertainties are still too large, for example the very high $\pt$ bins, are also excluded to optimize the tuning but we note that it is only applicable to a few regions of phase space of our observables. We explicitly turn off long-lived particle decays in our MC generation as all relevant observables have either been corrected for feed-down decays or unfolded to particle level distributions.

\section{New PYTHIA 8 Detroit Tune\label{sec:newtune}}

%\begin{figure}[!htbp]
%    \centering
%    \includegraphics[width=0.3\textwidth]{figs/mnprofile_MultipartonInteractionspT0Ref.pdf}
%    \includegraphics[width=0.3\textwidth]{figs/mnprofile_MultipartonInteractionsecmPow.pdf}
%    \includegraphics[width=0.3\textwidth]{figs/mnprofile_MultipartonInteractionscoreRadius.pdf}
%    \includegraphics[width=0.3\textwidth]{figs/mnprofile_MultipartonInteractionscoreFraction.pdf}
%    \includegraphics[width=0.3\textwidth]{figs/mnprofile_ColourReconnectionrange.pdf}
%    \caption{$\chi^{2}$ profiles for each tuning parameter in the vicinity of the global minimum (offset to have minimum at zero). The green shaded regions show $\Delta\chi^{2}=1$ from the best-fit values. }
%    \label{fig:min}
%\end{figure}
\begin{figure}[!htbp]
    \centering
    \includegraphics[width=0.9\textwidth]{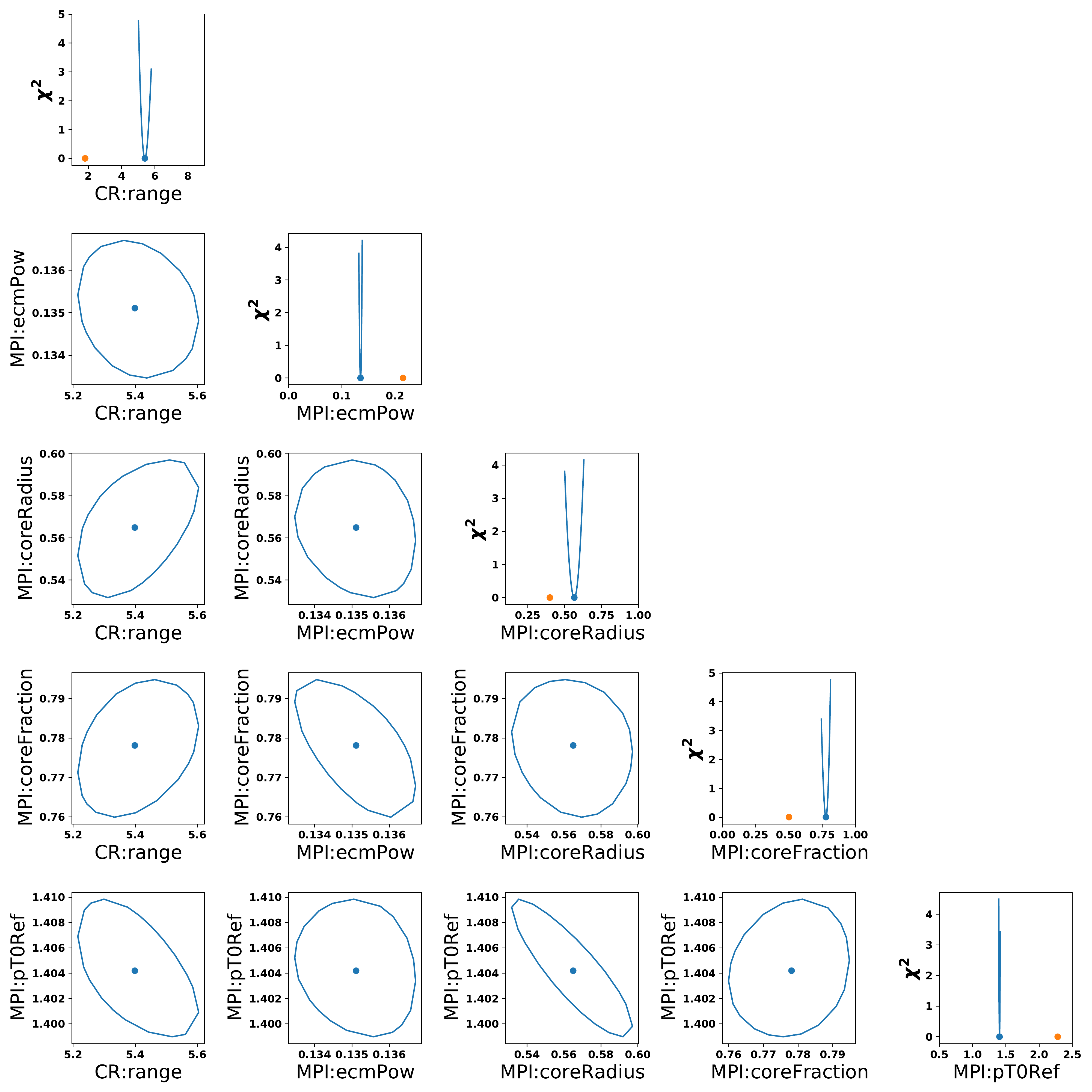}
    \caption{Minimum-subtracted $\chi^{2}$ profiles in the vicinity of the best fit value, shown in the right-most diagonal panels, and the one sigma correlation contours between all tuning parameters, shown on the off-diagonal panels. The single (orange) points along the diagonal panels show the default PYTHIA 8.303 values tuned at a reference energy of 7 TeV. Additionally, the $x$-axis scale in the diagonal panels show the allowed ranges used in the tuning procedure.}
    \label{fig:contour}
\end{figure}
Figure~\ref{fig:contour} show the $\chi^{2}$ profiles projected onto each tuning parameter in the vicinity of the global minimum (offset to have minimum at zero), and the one sigma correlation contours between all tuning parameters. The central values of the tuning parameters are tabulated in Table~\ref{tab:paramsnew}, and define the new 'Detroit' tune. In general the errors on the central values from Minuit are much smaller than the eigentune values quoted below, and are therefore not given. The $\chi^{2}$ per degrees of freedom ($n.d.f.$) for the best fit is $\chi^{2}/n.d.f.$ = 611/493. 

\begin{figure}[!htbp]
    \centering
    \includegraphics[width=0.45\textwidth]{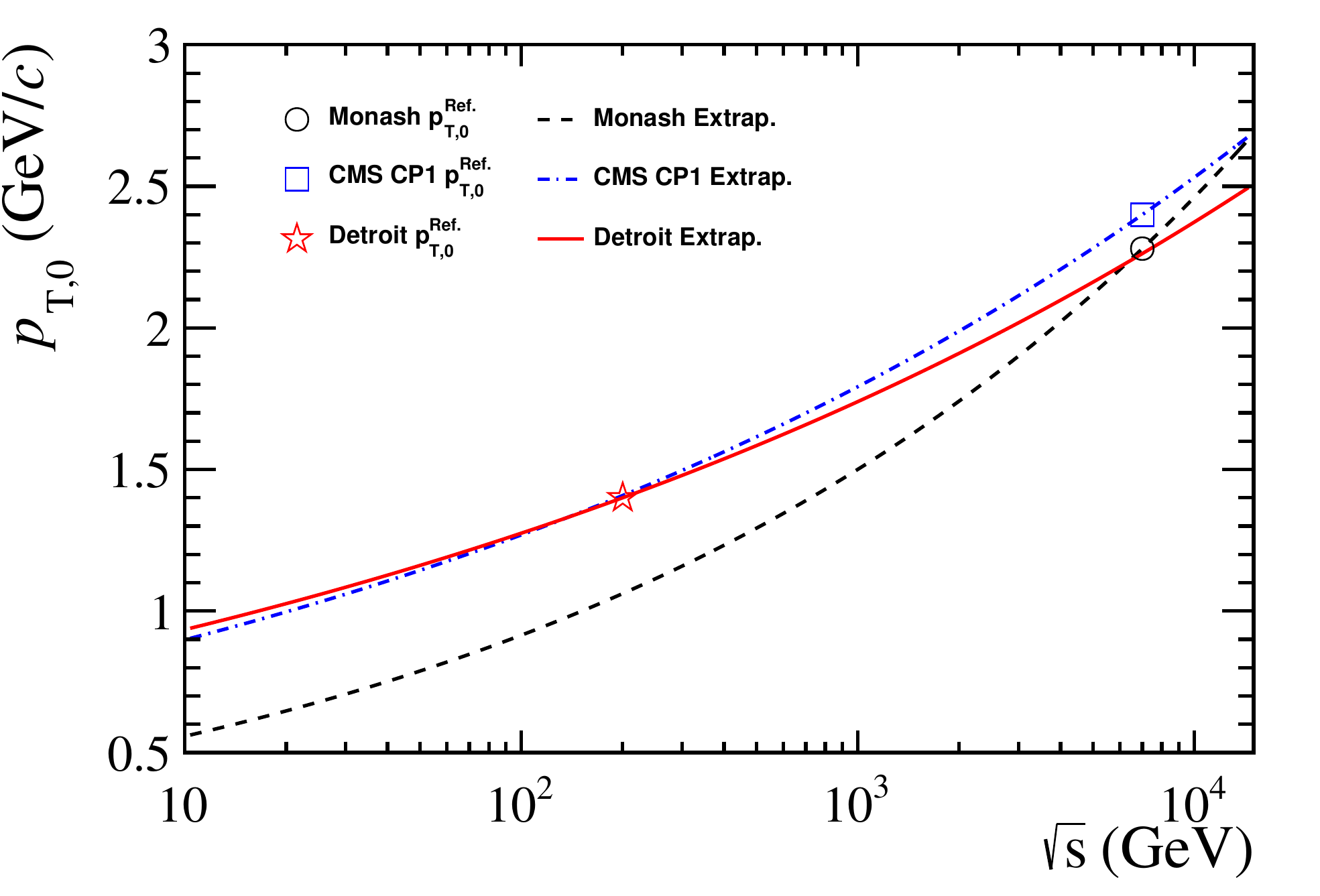}
    \caption{Values for $p_{T,0}^{Ref}$ for the Monash (open black circle)~\cite{Skands:2014pea}, CMS CP1 (open blue square)~\cite{CMS:2015wcf}, and Detroit (open red star) tunes. The energy extrapolations for the Monash, CMS CP1, and Detroit tunes are shown as the dashed black, dot-dashed blue, and solid red lines, respectively. }
    \label{fig:extrap}
\end{figure}

In Fig.~\ref{fig:extrap} we plot the values for $p_{T,0}^{Ref}$, at their respective reference energies, and their extrapolations for the Monash, CMS CP1~\cite{CMS:2015wcf}, and Detroit tunes. The extrapolations are determined as $p_{T,0}=p_{T,0}^{Ref}(\sqrt{s}/\sqrt{s}_{Ref.})^{ecmPow}$. We compare to the CMS CP1 tune as it followed a similar tuning strategy utilizing both LHC data and the CDF UE data at $\s$ = 900 and 1960 GeV, and the same PDFs. However, in the CMS CP1 tune the reference energy is kept at 7 TeV. The new $p_{T,0}^{Ref}$ value at a reference energy of $\s$ = 200 GeV is about 30\% larger compared to the $\s$ = 200 GeV extrapolated $p_{T,0}$ using the Monash values, and comparable to the extrapolated values in the CMS CP1 tune. However, in the latter comparison the values of $p_{T,0}$ diverge as you go up or down in collision energy. The value for the energy-depended extrapolation reduced by almost 40\% compared to the Monash tune, which significantly reduces how fast $p_{T,0}$ varies. Compared to the CMS CP1 tune, our new extrapolation parameter is roughly 10\% smaller. 

\begin{table}[!htbp]
	\caption{PYTHIA 8 tuned parameters. \label{tab:paramsnew}}
	\centering
	\begin{tabular}{c | c | c}
		\hline \hline
    Tuning Parameter & Default & Detroit  \\ \hline
    MultipartonInteractions:pT0Ref & 2.28 GeV &  1.40 GeV     \\    
    MultipartonInteractions:ecmPow & 0.215 & 0.135     \\   
    MultipartonInteractions:coreRadius & 0.4 & 0.56      \\ 
    MultipartonInteractions:coreFraction & 0.5 & 0.78      \\    
    ColourReconnection:range       & 1.8 & 5.4      \\   
		\hline \hline
	\end{tabular}
\end{table}

We observe that the proton overlap function parameters coreRadius and coreFraction, and color reconnection range to have slightly increased values in our tune compared to the default PYTHIA 8 and CMS CP1 values as presented in Table~\ref{tab:paramsnew}. We have also performed the tuning study with the default proton overlap shape function (MultipartonInteractions:bprofile=3) and find the description of the $\s$ = 300 GeV UE data to be inadequate, and had a global $\chi^{2}$ more than a factor of two larger with respect to the Detroit tune. 

\begin{table*}[!htbp]
	\caption{PYTHIA 8 tune parameter variations for each eigentune. \label{tab:eigen}}
	\centering
	\begin{tabular}{c|c c c c c c c c c c}
		\hline \hline
    Tuning Parameter & 1+ & 1- & 2+ & 2- & 3+ & 3- & 4+ & 4- & 5+ & 5-    \\ \hline
    MultipartonInteractions:pT0Ref (GeV)&  1.37  & 1.43&  1.38  & 1.42   &   1.44 &1.37    &  1.41  & 1.40  & 1.40 &  1.41  \\    
    MultipartonInteractions:ecmPow      & 0.132   &  0.138 & 0.135& 0.135 & 0.119  & 0.150   & 0.145 & 0.126 &0.148 & 0.125 \\  
    MultipartonInteractions:coreRadius  &  0.74  &   0.41 &  0.77  & 0.41   &  0.57  & 0.56   &  0.57  & 0.56  & 0.51 & 0.60 \\
    MultipartonInteractions:coreFraction&  0.84  & 0.72   & 0.72   & 0.82   &  0.78  &  0.78  & 0.78   &  0.78 &0.60 &0.90  \\  
    ColourReconnection:range            &  7.50  &  3.61  & 5.38   & 5.41   & 5.40   &  5.40  & 5.40   & 5.40 & 5.41& 5.40  \\  	\hline \hline 

		\end{tabular}
\end{table*}

We also provide a set of 'eigentunes' that quantify the MC errors that may be used for systematic studies. This was done using the Professor package which diagonalizes the covariance matrix at the best fit point, and provides values for the tuned parameters that correspond to deviations along the principle directions for a fixed tolerance $\Delta \chi^{2}$. For our analysis, we choose a heuristic $\Delta \chi^{2}$=$n.d.f./2$ as done in Ref.~\cite{TheATLAScollaboration:2014rfk}. Table~\ref{tab:eigen} tabulates the parameter values for each eigentune.  

\section{Comparisons with Mid-Rapidity Data\label{sec:comp}}

\begin{figure}[!htbp]
    \centering
    \includegraphics[width=0.3\textwidth]{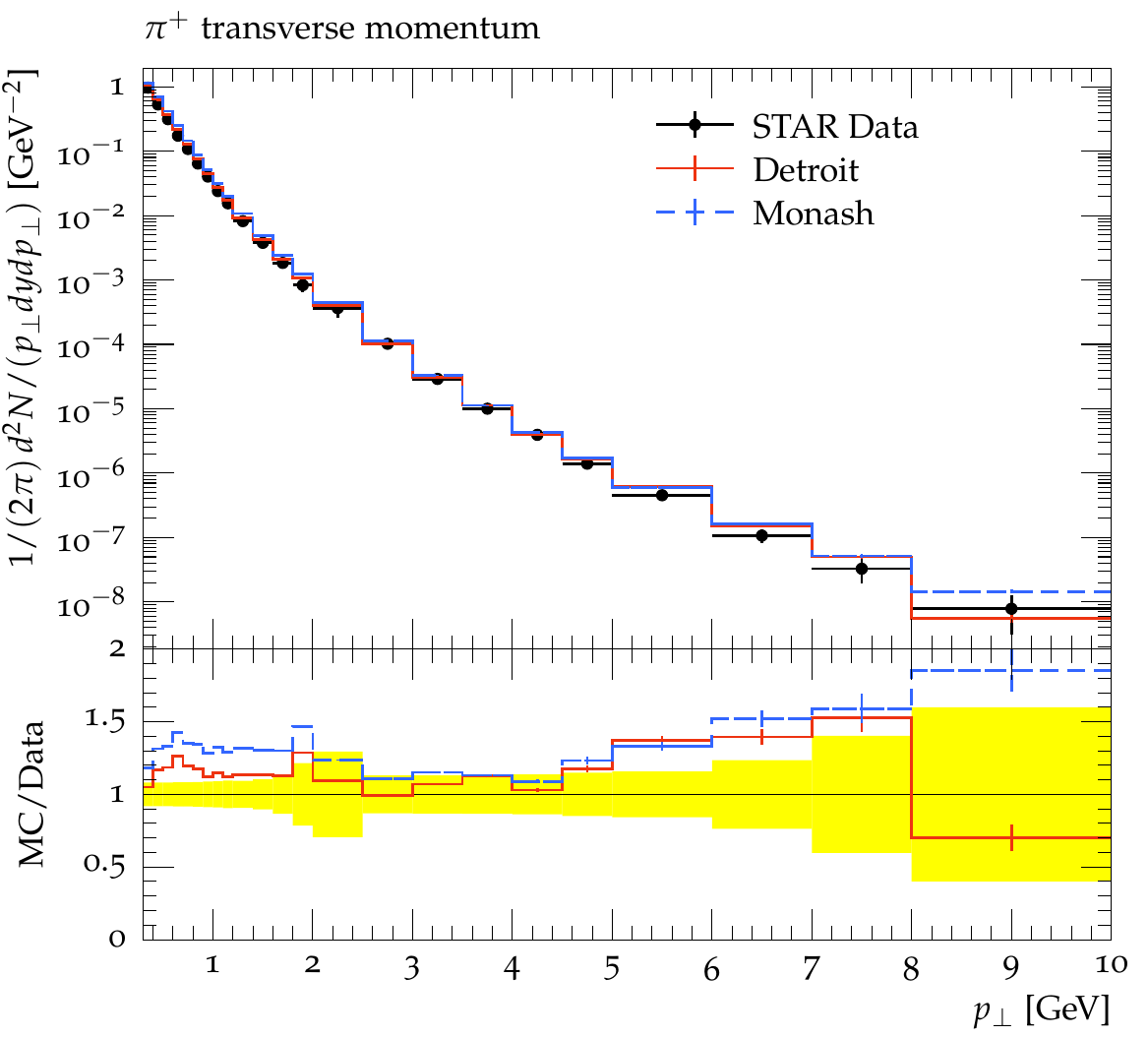}
    \includegraphics[width=0.3\textwidth]{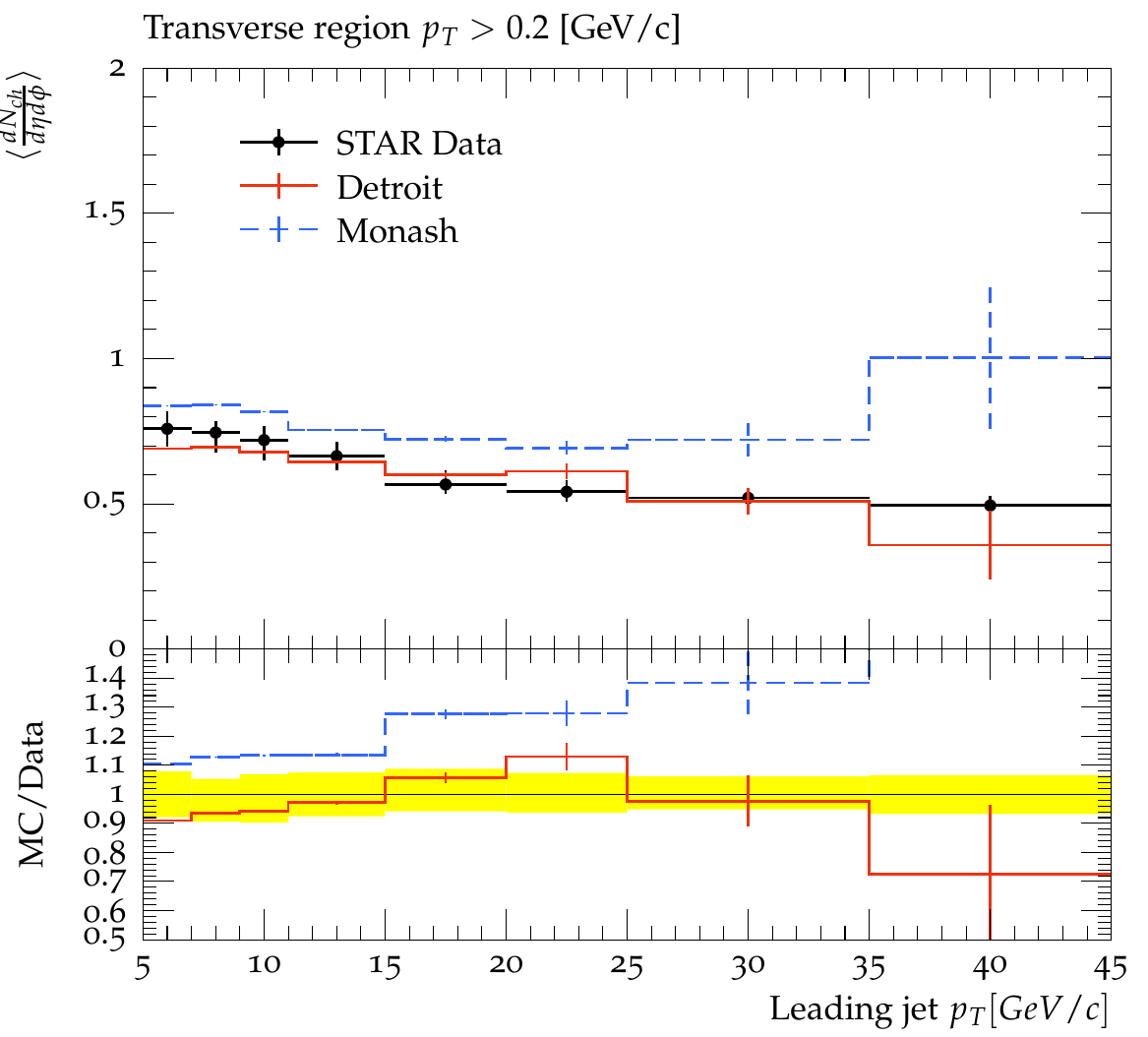}
    \includegraphics[width=0.3\textwidth]{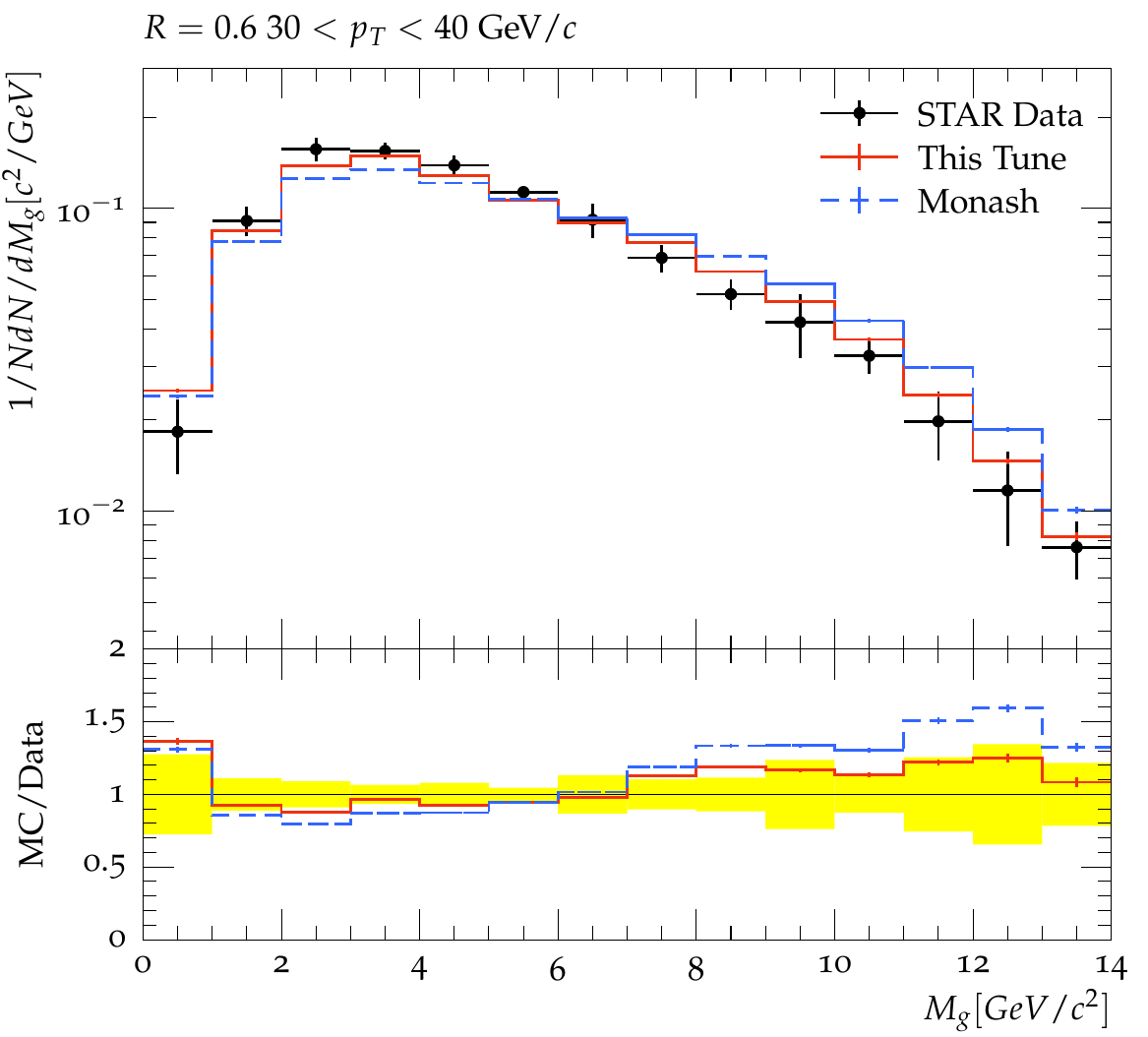}     
    \caption{Comparison of the default (blue dashed) and Detroit PYTHIA 8 tunes (red solid) with mid-rapidity $\pi^{+}$ cross sections as a function of $p_{T}$ (left)~\cite{ADAMS2006161}, UE multiplicity as a function of leading jet $\pt$ (middle)~\cite{PhysRevD.101.052004}, and the SoftDrop groomed jet mass (right)~\cite{starcollaboration2021invariant} in $p$+$p$ collisions at $\s$ = 200 GeV measured by the STAR experiment. The bottom panels in each figure show the ratios of the Monte Carlo predictions with respect to the data and the yellow shaded region shows the data uncertainties.}
    \label{fig:datacomp}
\end{figure}

Figure~\ref{fig:datacomp} shows the comparison of the PYTHIA 8 Monash and Detroit tunes for a representative sample of measurements in data. All others are shown in Appendix~\ref{app:a}. The mid-rapidity $\pi^{+}$ cross section on the left, the average UE transverse charge particle multiplicity in the middle, and the groomed jet mass distribution on the right for $\s$ = 200 GeV $p$+$p$ collisions. We find, in general, the Detroit tune provides a better match to the data than the Monash tune, in particular the underlying event charged particle multiplicity. It is especially relevant to highlight that reducing the $\pi$ cross section at low momenta directly translates to a significantly better description of the UE multiplicity and the jet mass, especially in the tails of the distribution. 

In $\s$ = 200 GeV $p$+$p$ collisions, the $\pi^{-}$ spectrum is well described by the Detroit tune across the entire measured $p_{T}$ range. The $\pi^{+}$ spectrum is consistent except still being slightly over predicted above $p_{T}$ = 5 GeV/$c$ by about 10\% considering experimental uncertainties. The charged particle multiplicities in the toward, away, and transverse regions are well described by the Detroit tune across all leading jet $p_{T}$. The average charge particle $p_{T}$ versus leading jet $p_{T}$ for the toward, away, and transverse regions changed only slightly in the Detroit tune with respect to the Monash tune. In the transverse region the description of the data is slightly improved and in the toward and away regions slightly degraded by a few percent. For the average $p_{T}$ for tracks with $p_{T}>$ 0.5 GeV/$c$ in the transverse region, we observe both the Detroit and Monash tunes undershoot the data at low leading jet $p_{T}$ by around 8\% and improve at higher leading jet $p_{T}$. The Drell-Yan di-muon data is well described by the Detroit tune where as the Monash tune slightly under predicted the data. In all the jet-substructure observables we see an improved agreement with the data using the Detroit tuned PYTHIA, but observe some persistent discrepancies in the tails of the distributions.

We observe that the Detroit tune is able to describe the CDF charge particle multiplicities in all $\sqrt{s}=$ 300, 900, and 1960 GeV center-of-mass energies, as shown in Appendix~\ref{app:a}. The charged particle $p_{T}$ sum is well described at $\sqrt{s}=$ 1960 GeV. In the $\sqrt{s}=$ 300 and 900 GeV charged particle $p_{T}$ sum observables the Detroit tune slightly under predicts the data in the leading particle $p_{T}$ range of around 4 to 8 GeV/$c$, but is less than 10\% considering experimental uncertainties.

\begin{figure}[!htbp]
    \centering
    \includegraphics[width=0.35\textwidth]{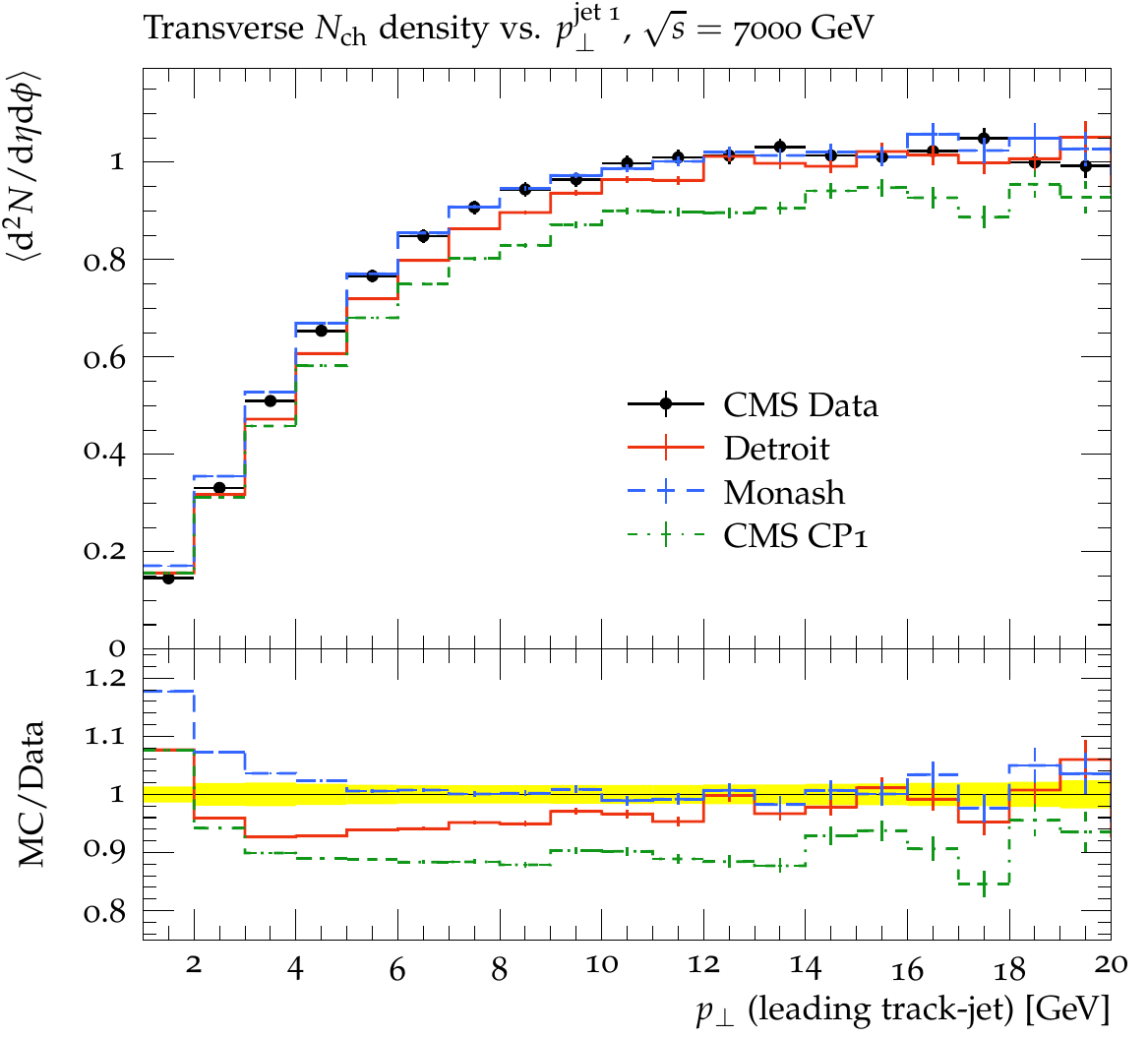}
    \includegraphics[width=0.35\textwidth]{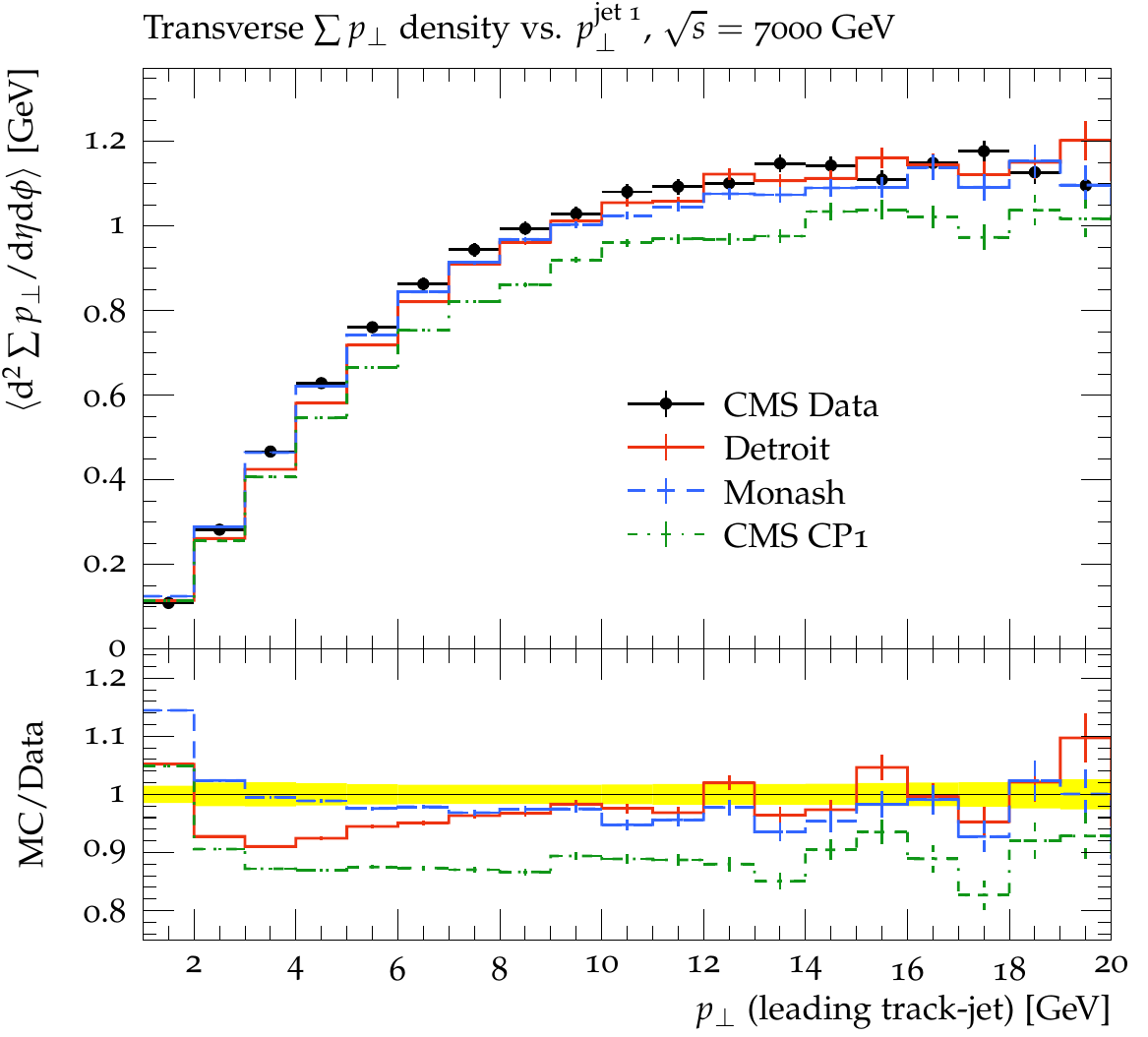}
    \caption{Underlying event observable in the transverse region as a function of leading track-jet $p_{T}$ from the CMS measurement in proton-proton collisions at $\s$ = 7 TeV~\cite{Chatrchyan2011}. The left shows the charge particle multiplicity and the right shows the $p_{T}$ sum. The bottom panels in each figure show the ratios of the Monte Carlo predictions with respect to the data and the yellow shaded region shows the data uncertainties.}
    \label{fig:cms1}
\end{figure}

\begin{figure}[!htbp]
    \centering
    \includegraphics[width=0.35\textwidth]{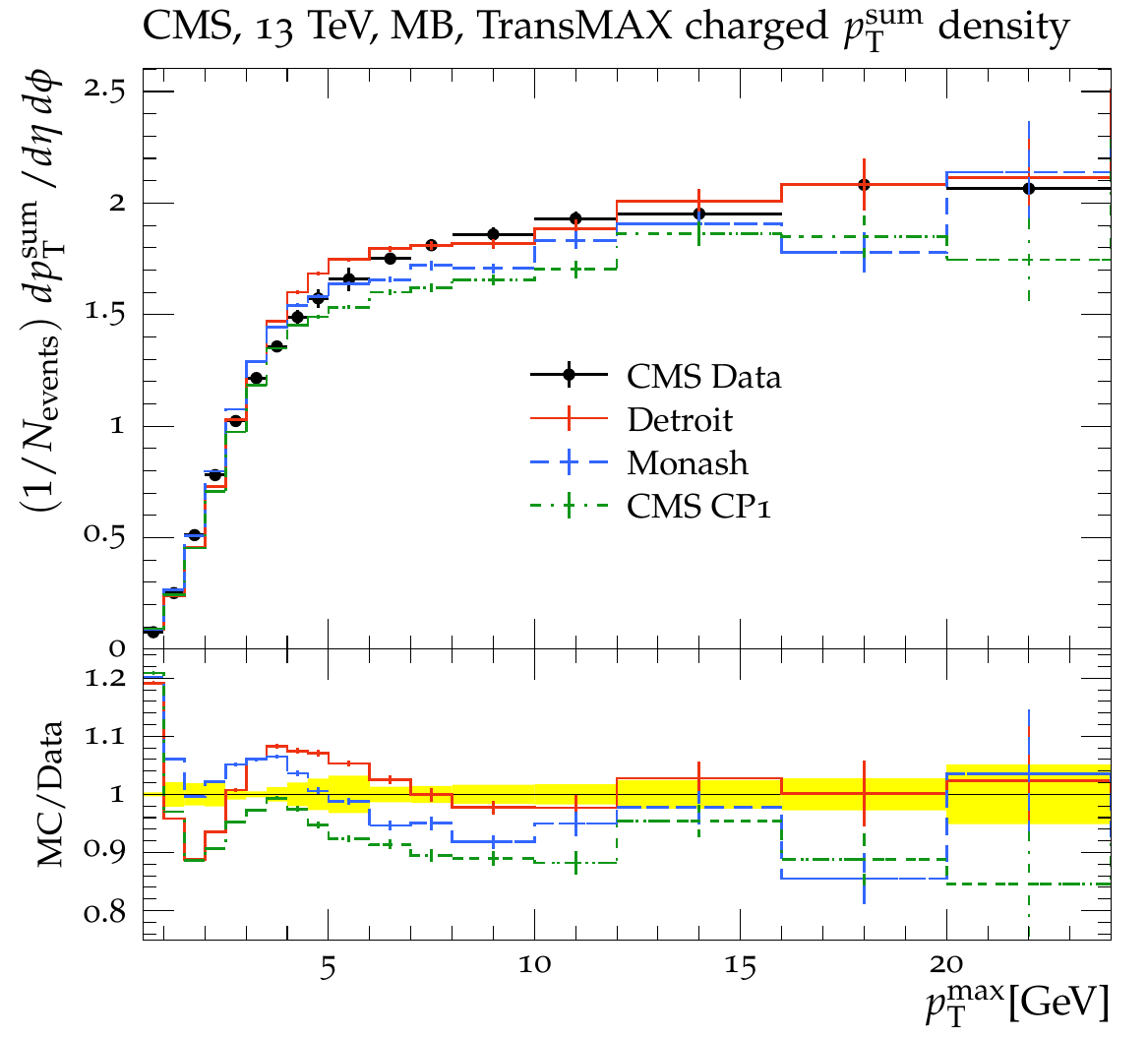}
    \includegraphics[width=0.35\textwidth]{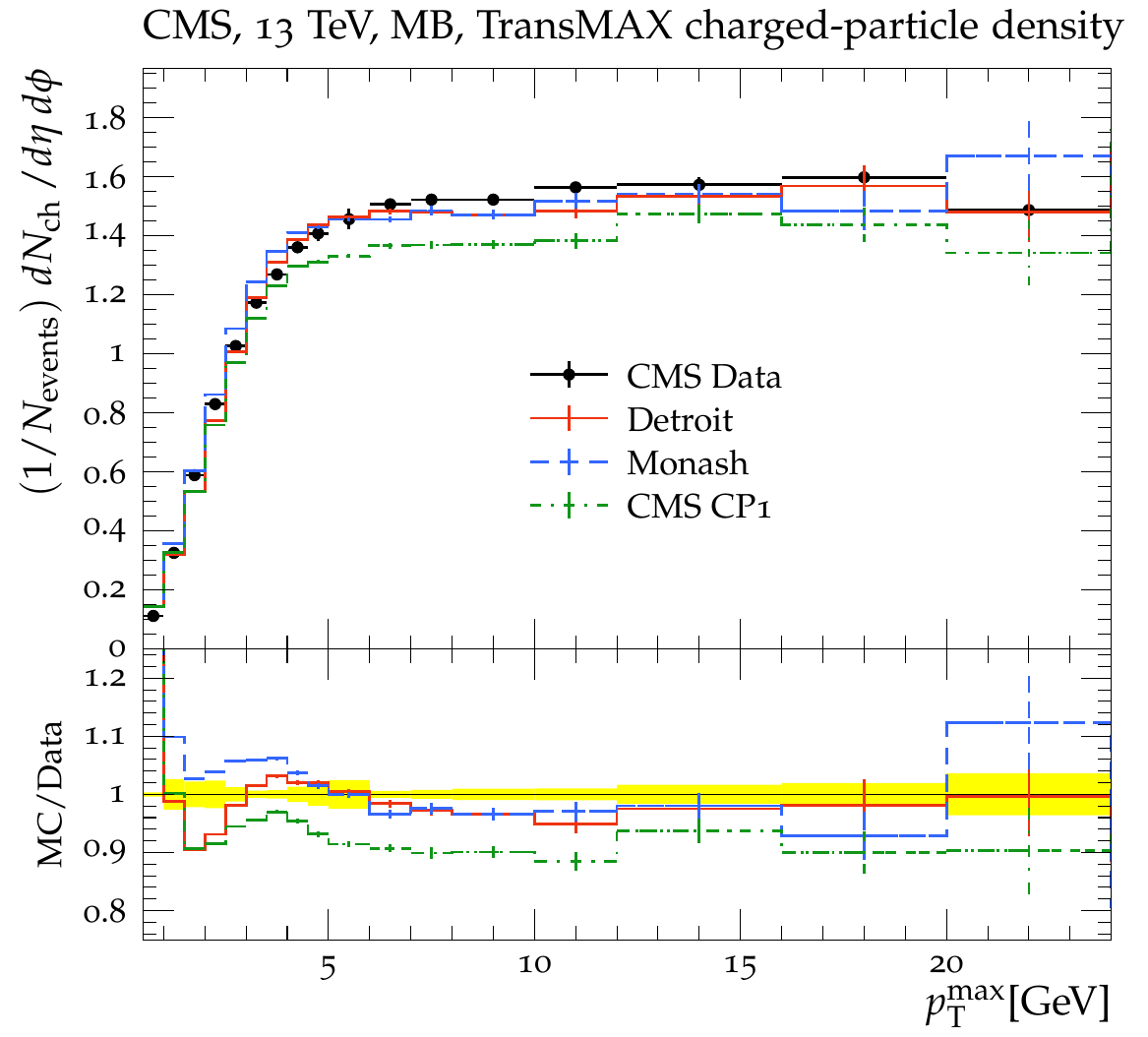}
    \includegraphics[width=0.35\textwidth]{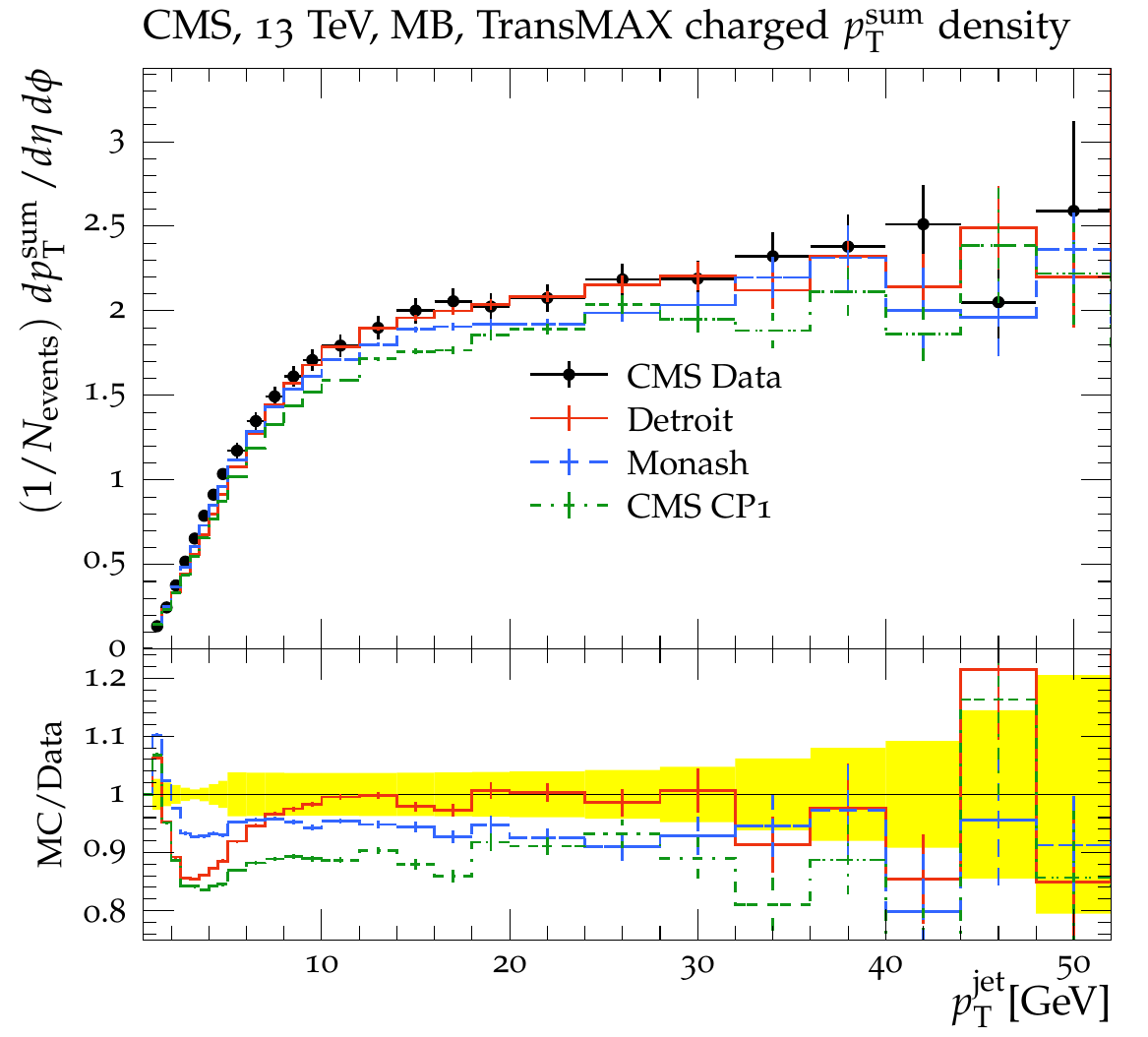}
    \includegraphics[width=0.35\textwidth]{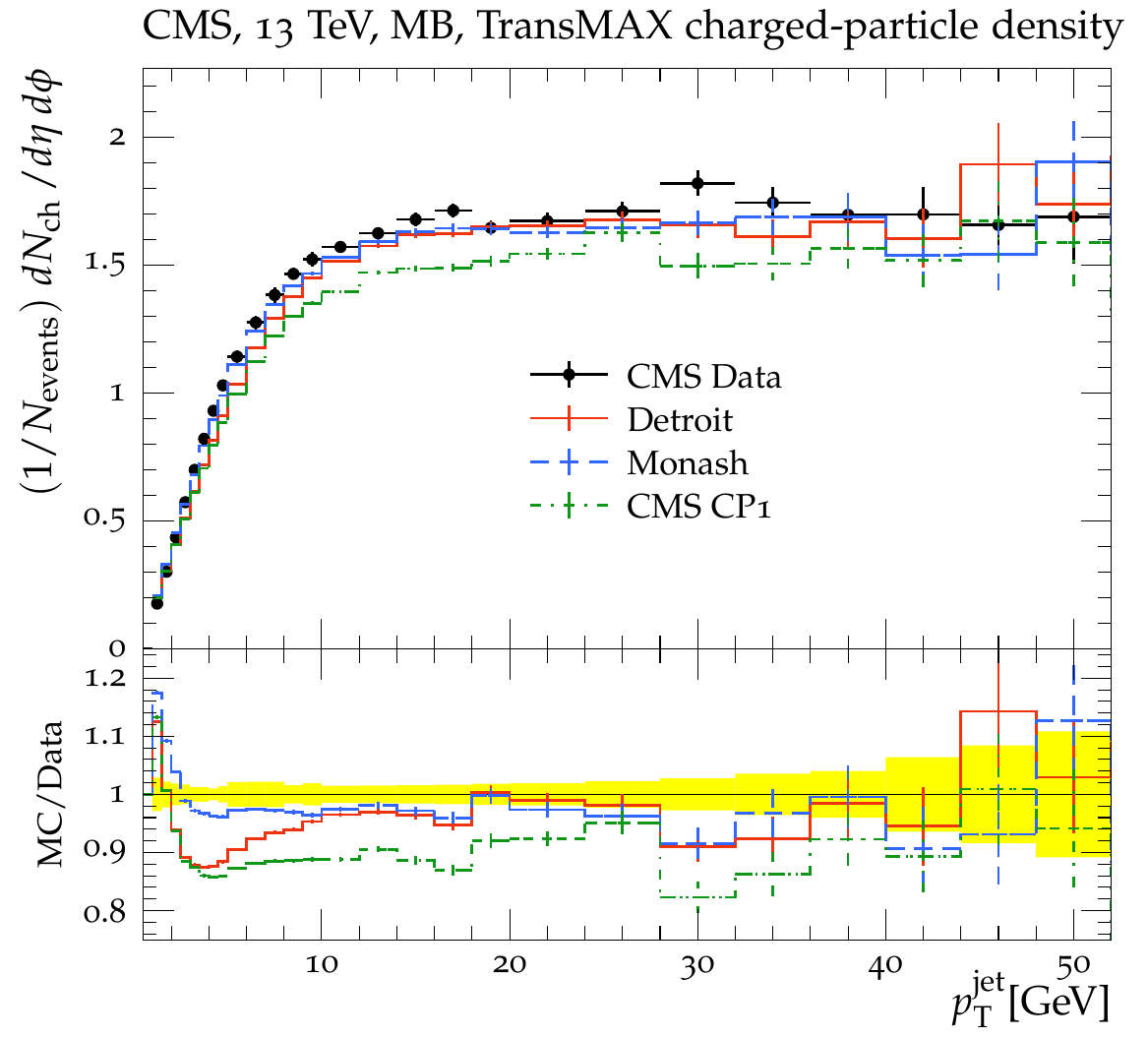}
    \caption{The $p_{T}$ sum (left) and charge particle multiplicity (right) as a function of leading track (top) and jet (bottom) $p_{T}$ in the transMax region from the CMS measurement in proton-proton collisions at $\s$ = 13 TeV~\cite{CMS:2015zev}. The bottom panels in each figure show the ratios of the Monte Carlo predictions with respect to the data and the yellow shaded region shows the data uncertainties.}
    \label{fig:cms2}
\end{figure}

We compare our tune with the Monash and CMS CP1 tunes to underlying event observables measured by the CMS experiment in data from 7~\cite{Chatrchyan2011} and 13 TeV~\cite{CMS:2015zev} $p$+$p$ collisions in Figs.~\ref{fig:cms1} and~\ref{fig:cms2}, respectively (additional 13 TeV comparisons in Appendix~\ref{app:a}). We note for all these comparisons, we have explicitly turned off long-lived decays in the MC generation. In the 7 TeV data, the Monash tune is able to describe the data better than the Detroit tune and the CMS CP1 tune. At higher leading track-jet $p_{T}$, the Detroit tune is consistent with the data and the CMS CP1 tune slightly under predicts the data. At 13 TeV, our tune is able to describe the data above a leading track or jet $p_{T}$ of 5 and 10 GeV, respectively. The Monash tune is able to describe the charged particle density in the same regions, but under predicts the $p_{T}$ sum. The CMS CP1 tune under predicts both. We note that in the low $p_{T}$ regions, the predictions from our tune vary more in shape with respect to the Monash tune, and is due to the proton shape function used.

\section{Comparisons at forward rapidity\label{sec:forward}}

\begin{figure*}[!htbp]
    \centering
    \includegraphics[width=0.35\textwidth]{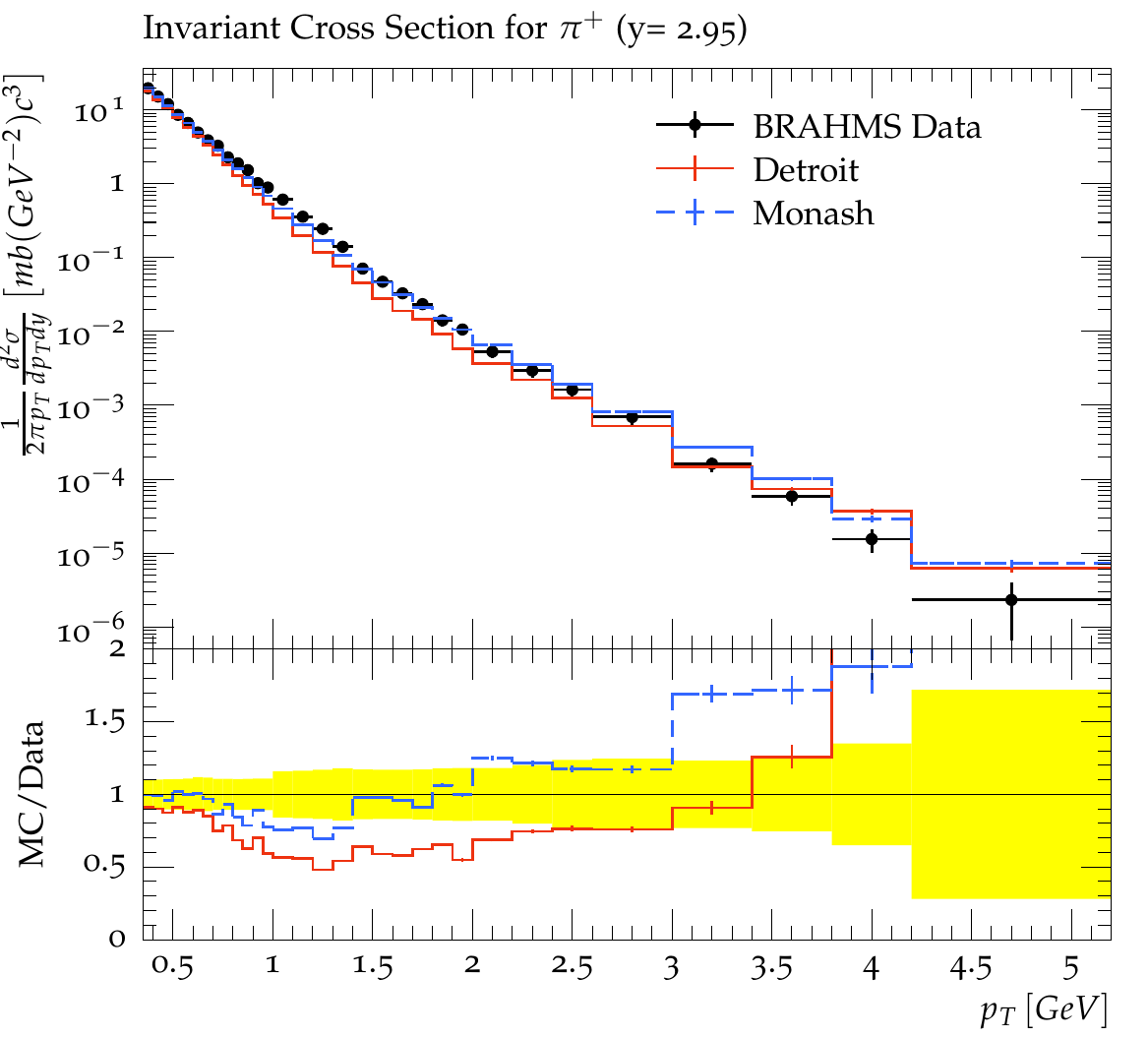}
    \includegraphics[width=0.35\textwidth]{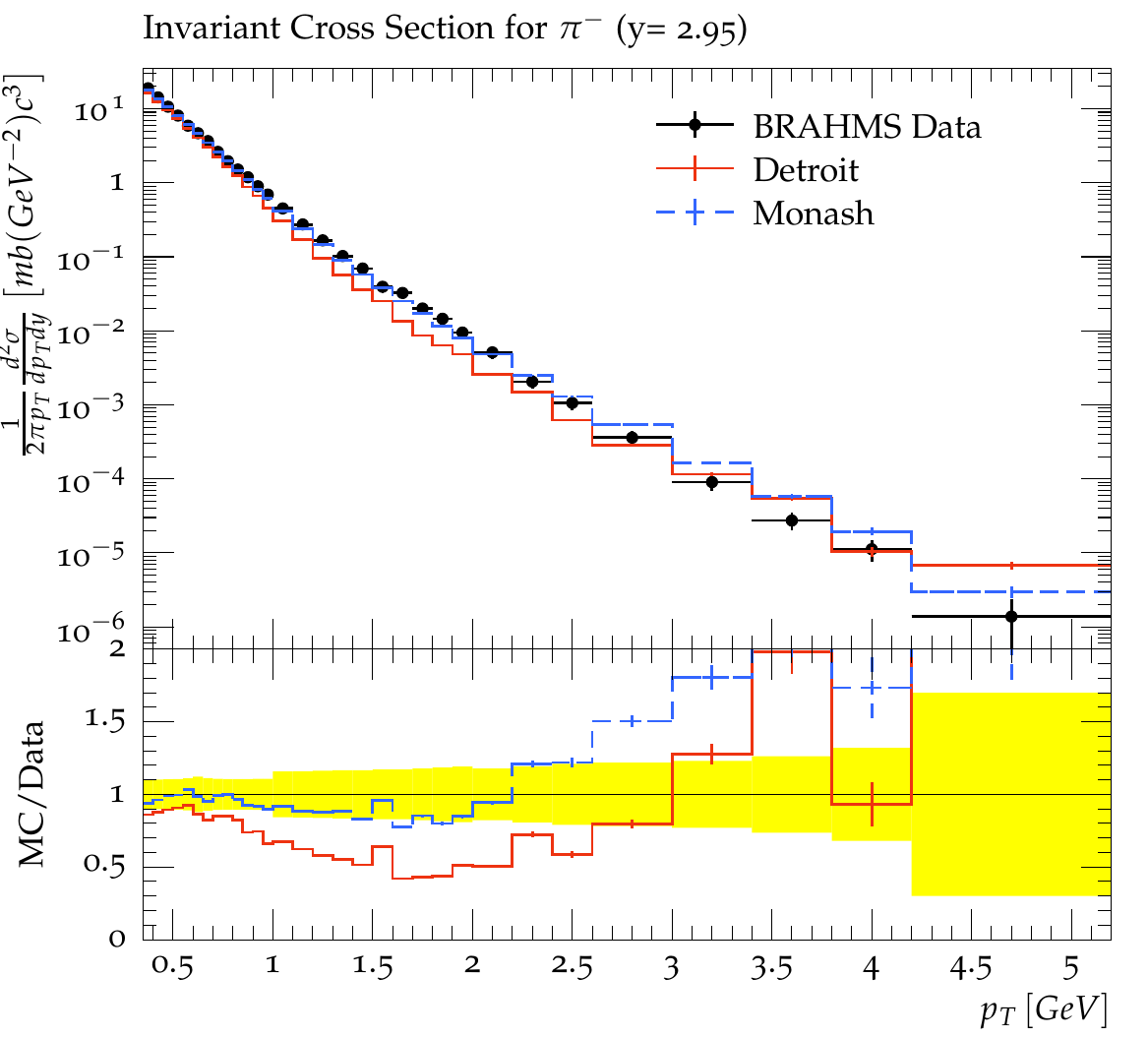}
    \includegraphics[width=0.35\textwidth]{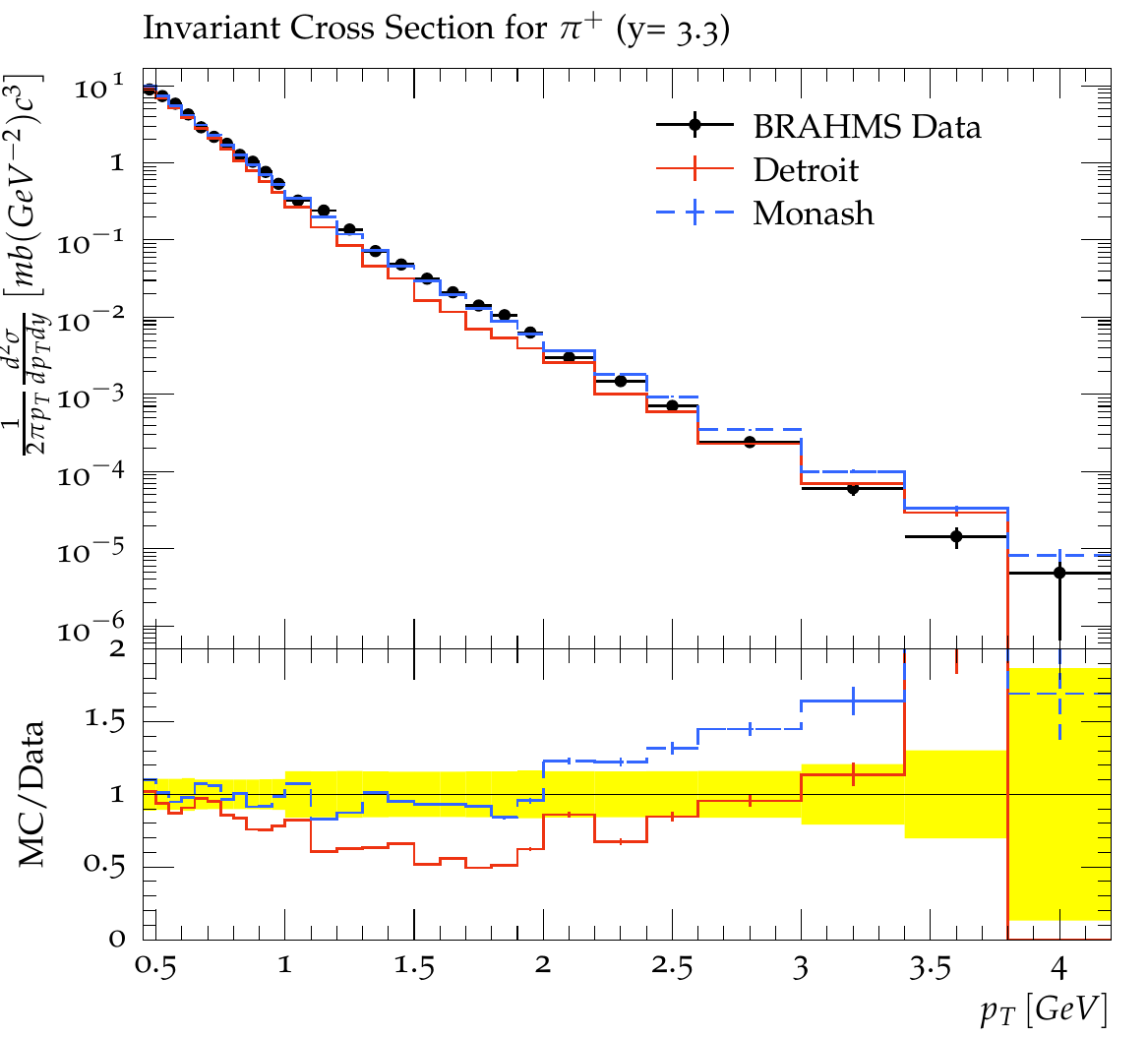}
    \includegraphics[width=0.35\textwidth]{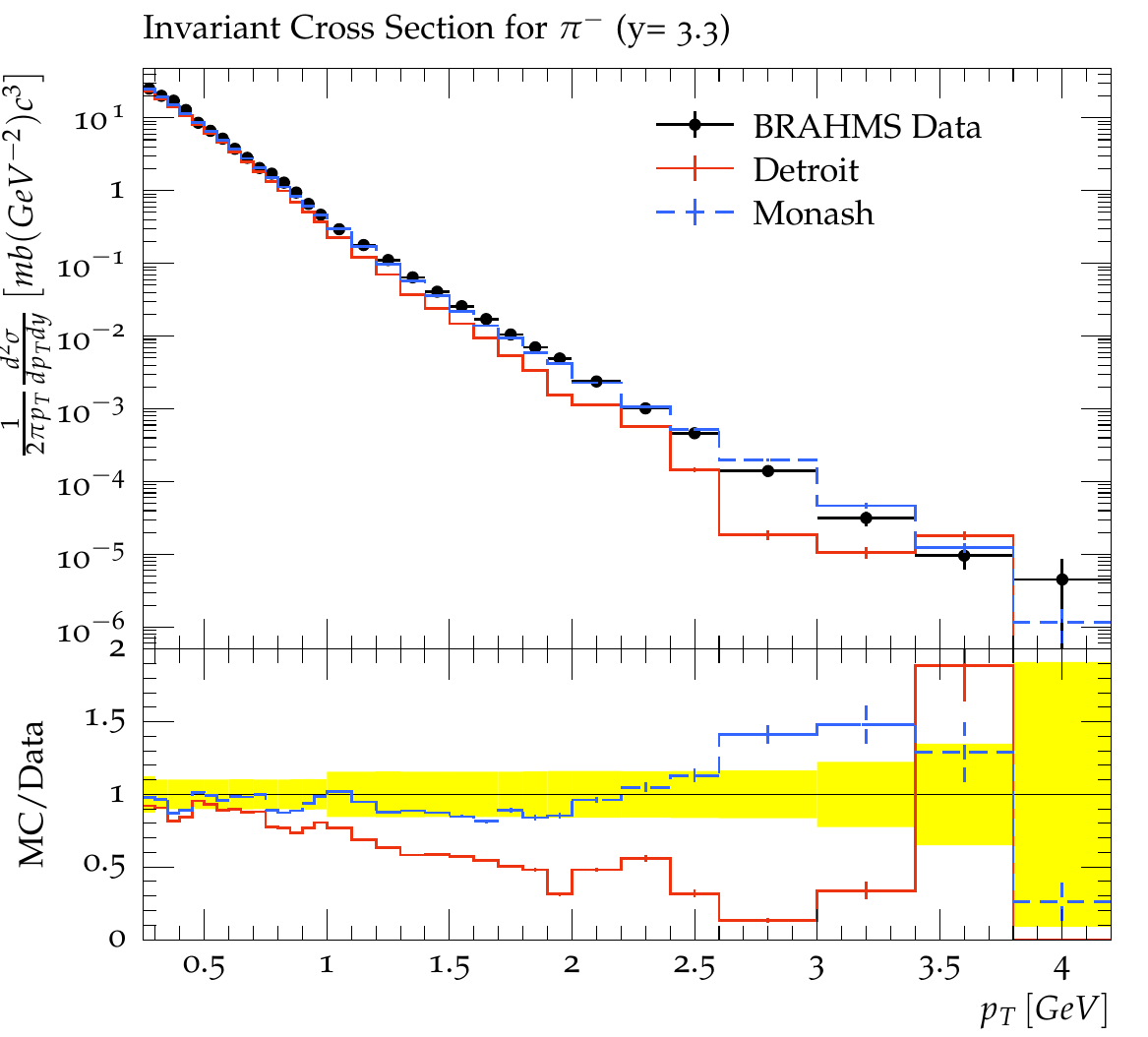}    
    \caption{$\pi^{\pm}$ cross sections at forward rapidity in $p$+$p$ collisions at $\s$ = 200 GeV measured by the BRAHMS experiment~\cite{PhysRevLett.98.252001} compared to the  default (blue dashed) and Detroit (red solid) PYTHIA 8 tunes. The top two figures show $\pi^{+}$ and $\pi^{-}$, at $y$ = 2.95. The bottom two show the same except for $y$ = 3.3. The bottom panels in each figure show the ratios of the Monte Carlo predictions with respect to the data and the yellow shaded region shows the data uncertainties.  }
    \label{fig:brahms}
\end{figure*}

\begin{figure}[!htbp]
    \centering
    \includegraphics[width=0.4\textwidth]{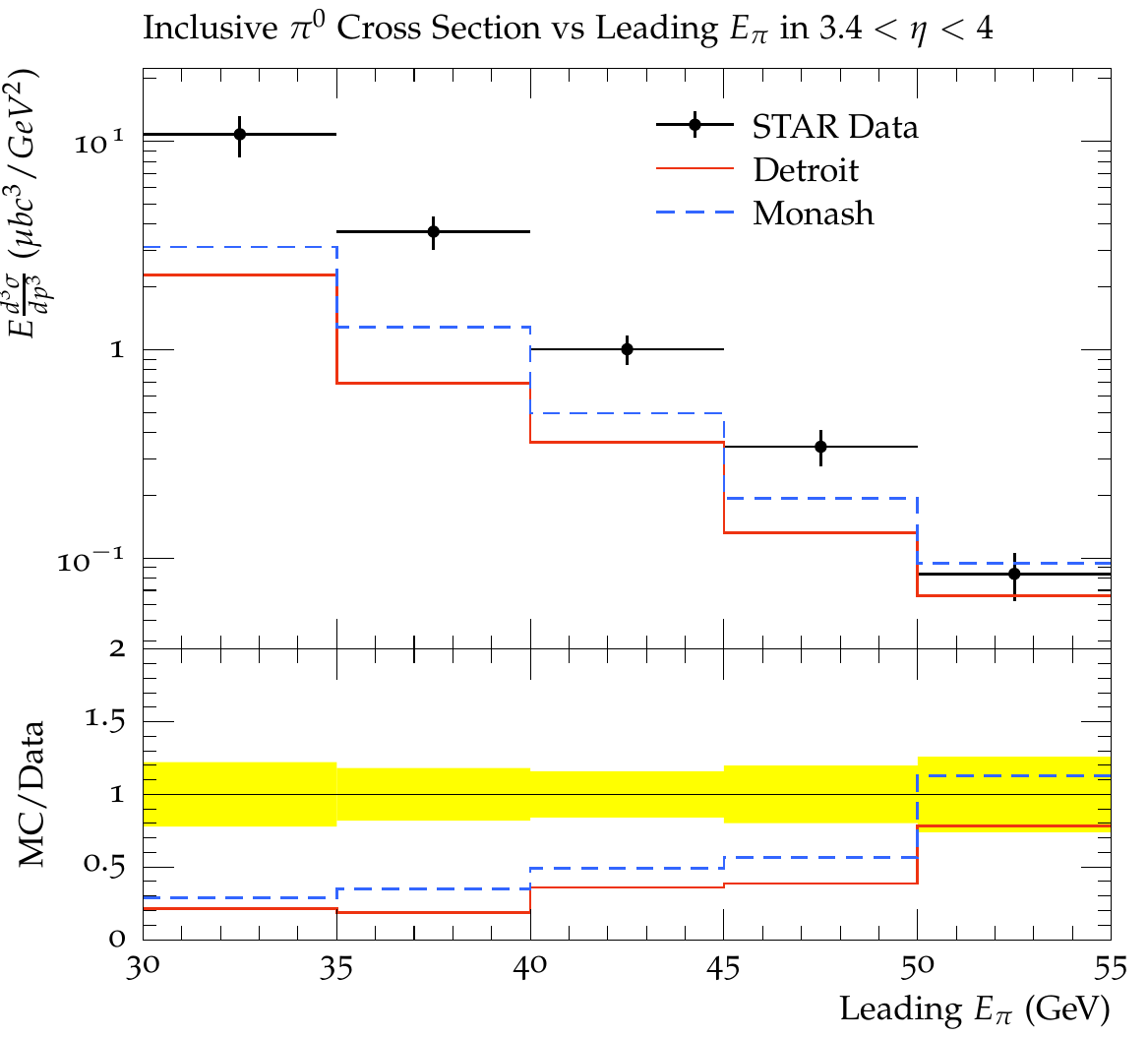}
    \caption{Inclusive $\pi^{0}$ cross sections versus leading $\pi^{0}$ energy at forward rapidity in $p$+$p$ collisions at $\s$ = 200 GeV measured by the STAR experiment~\cite{PhysRevLett.92.171801} compared to the  default (blue dashed) and Detroit (red solid) PYTHIA 8 tunes. The bottom panels in each figure show the ratios of the Monte Carlo predictions with respect to the data and the yellow shaded region shows the data uncertainties.  }
    \label{fig:starforward}
\end{figure}

We compare the Detroit and Monash PYTHIA 8 tunes to the measured $\pi^{\pm}$ cross sections at rapidity $y$ = 2.95 and 3.3 measured by the BRAHMS experiment~\cite{PhysRevLett.98.252001} and the inclusive $\pi^{0}$ cross sections in 3.4$<\eta<$4.0 measured by the STAR experiment~\cite{PhysRevLett.92.171801} in $\s$ = 200 GeV $p$+$p$ collisions. As shown in Fig.~\ref{fig:brahms} and~\ref{fig:starforward} for BRAHMS and STAR data, respectively, both the Monash and Detroit tunes undershoot the data in the measured low and mid $p_{T}$/$E_{\pi}$ ranges, and are consistent or exceed the data at high $p_{T}$/$E_{\pi}$. In the case of the Detroit tune, this discrepancy is persistent even considering the envelope covered by the eigentunes in Table~\ref{tab:eigen}. An interesting observation is that the Detroit tune does worse compared to the standard tune. We have attempted simultaneous tuning of mid- and forward-rapidity data, but are unable to recover the agreement seen for the mid-rapidity tune. The forward rapidity data favors larger values of $p_{T,0}^{Ref.}$ compared to the Monash value to induce a better MC agreement with the data. We additionally have expanded our simultaneous tuning exercise to include the initial state radiation parameters $\alpha_{s}$, intrinsic $k_{T}$, and (SpaceShower) $p_{T,0}^{Ref.}$, and utilizing a different proton shape parameterization (MultipartonInteractions:bProfile=3) with its associated parameter. Even with the inclusion of these additional tuning parameters and an enhanced tunable phase-space, we are still unable to reach a satisfactory agreement with both mid- and forward-rapidity data. 

Having a PYTHIA tune that is able to simultaneously describe the physics at both mid- and forward-rapidity regions will be essential for the upcoming RHIC program. The STAR experiment has installed a new detector system in the forward region (2.5$<\eta<$4) for the 2022 Run~\cite{starforward}. Electromagnetic and hadronic calorimeters, as well as silicon trackers, will allow for full jet reconstruction for the first time in the far forward region. This will open up significantly enhanced opportunities for forward physics and mid-forward rapidity correlation studies with respect to previous data. Additionally, with the Electron-Ion collider program on the horizon spanning a broad kinematic range and excellent detector capabilities across rapidities, event generators will need to describe physics at the level of the experimental precision in order to advance the discovery potential. A forward tuning exercise that will aim to address these points will be the focus of future studies. 

\section{Conclusion\label{sec:conclusion}}

This publication represents the first UE tuning exercise of the PYTHIA 8 generator utilizing data in proton-proton collisions at center-of-mass energies down to 200 GeV. We have provided a new UE tune, which we name the `Detroit' tune, and a set of eigentunes for the PYTHIA 8 event generator that utilizes data from $\sqrt{s}$= 200, 300, 900, and 1960 GeV hadron-hadron collisions. We observe that the default Monash tune, that was tuned to data at LHC energies, significantly deviates from STAR, PHENIX, and CDF measurements. This is due to the energy-scaling of the PYTHIA $p_{T,0}$ regularization parameter not being appropriate. Our re-tuning of this parameter, and others related to the MPI, show a significant improvement in the simultaneous description of the data across all center-of-mass energies that are studied. We additionally compared our tune at LHC energies and find agreement with the UE data at higher $p_{T}$. This agreement is either comparable or better than the default Monash tune. At low $p_{T}$ there are large deviations with respect to the data, and are driven by the different proton shape function used. 

We advocate for using this tune for PYTHIA 8 studies at both current (STAR) and future (sPHENIX) RHIC experiments for the higher statistics running period planned for 2022-2025. Also, there are possibilities for sequential re-tunings of PYTHIA 8 as more data becomes available at RHIC energies, particularly at forward rapidities, but also to additionally improve jet-substructure observables. Therefore, our tune serves as a natural starting point for these future exercises. Our tuning study encourages the authors of the respective MC models to take into account data at a variety of collision energies towards a universal parameterizations of physics leading to improvements across varied kinematics.  

\section*{Acknowledgements}\label{sec:ack}
We would like the STAR collaboration and its management for initiating such a tuning exercise and their support and comments during the tuning process has indelibly contributed to its successful completion. We also like to thank Elke-Caroline Aschenauer and Carl Gagliardi in particular for input during the tuning. Special thanks to Flemming Videbaek for access to BRAHMS data. We express our gratitude to Christian Bierlich for useful discussions throughout this work. We thank the RCF facility at Brookhaven National Laboratory and the Wayne State University high performance computing center for the computational resources used for this work. This work was supported in part by the Office of Nuclear Physics within the U.S. DOE Office of Science, the U.S. National Science Foundation, and the National Natural Science Foundation of China.

\bibliography{bib}

\appendix

\section{Comparisons of the Detroit PYTHIA 8 tune with experimental data}\label{app:a}

\begin{figure*}[!htbp]
    \centering
    \includegraphics[width=0.27\textwidth]{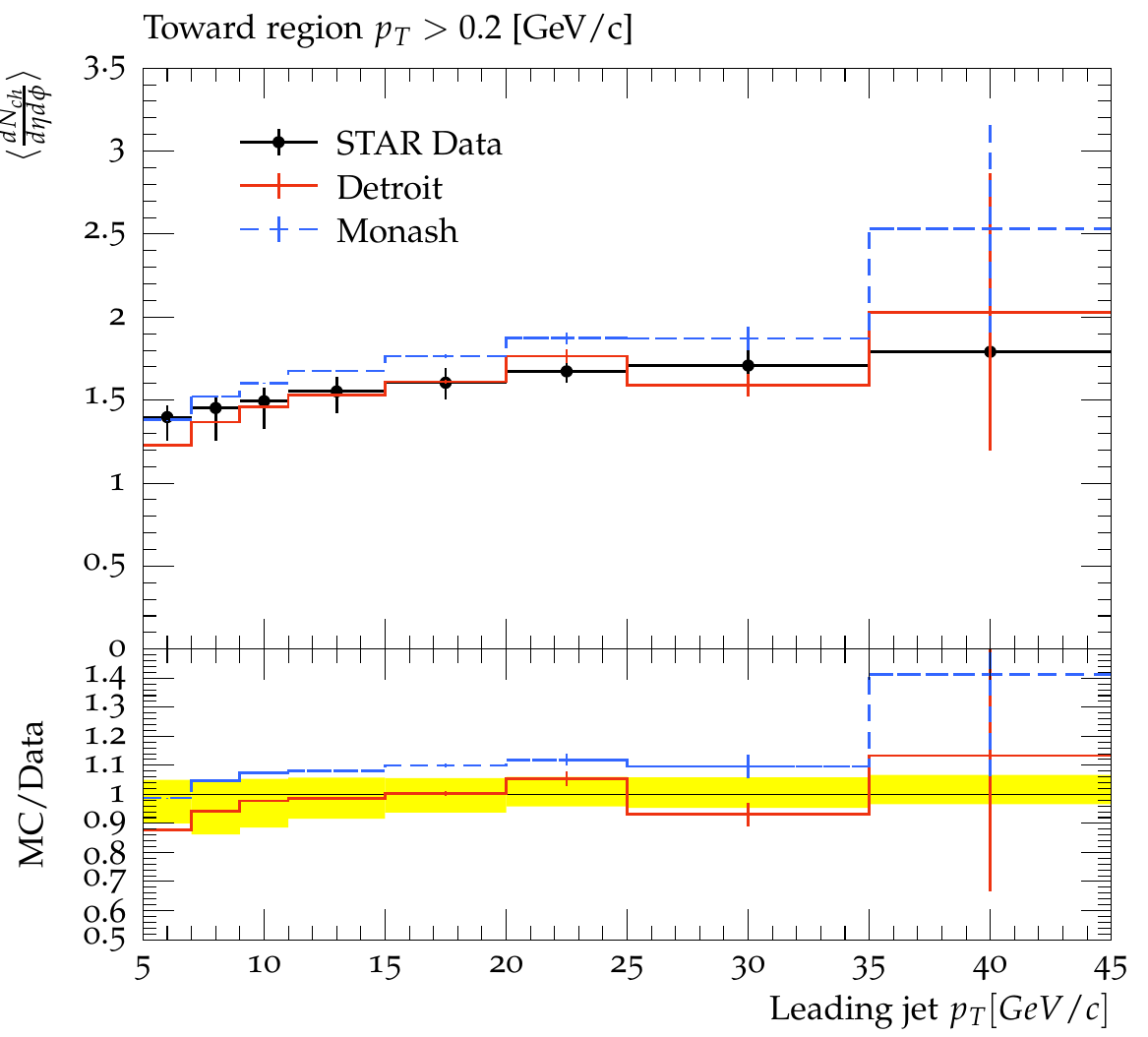}
    \includegraphics[width=0.27\textwidth]{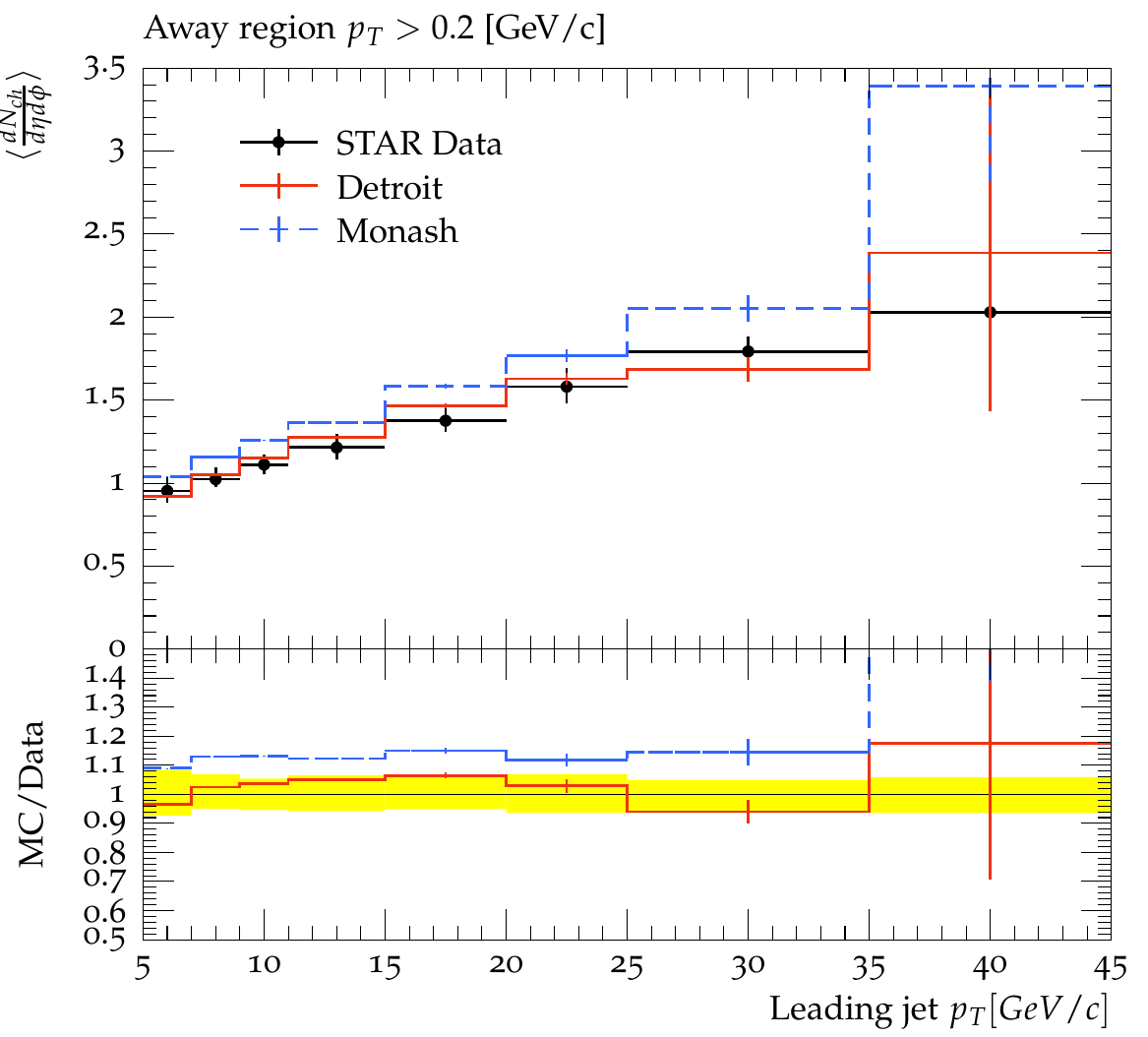} 
    \\
    \includegraphics[width=0.27\textwidth]{figs/STAR_2019_I1771348_d01-x01-y03.pdf}
    \includegraphics[width=0.27\textwidth]{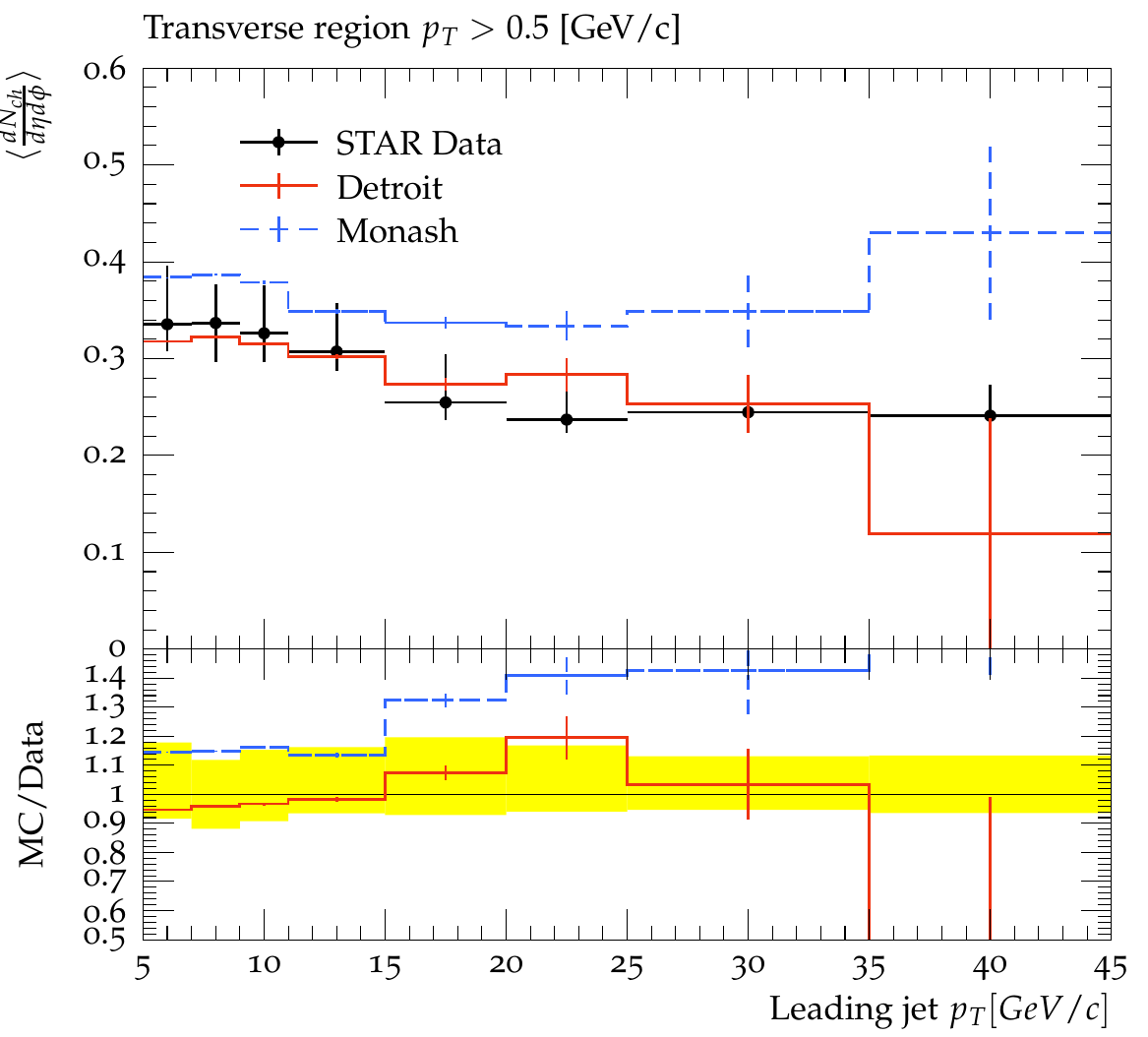}    
    \caption{Average charged particle multiplicity as a function of leading jet in $p$+$p$ collisions at $\s$ = 200 GeV measured by the STAR experiment~\cite{PhysRevD.101.052004}. The top left and right, and bottom left show for the toward, transverse, and away regions, respectively, for charged particles with $p_{T}>$ 0.2 GeV/$c$. The bottom right shows the transverse region for charged particles with $p_{T}>$ 0.5 GeV/$c$. The bottom panels in each figure show the ratios of the Monte Carlo predictions with respect to the data and the yellow shaded region shows the data uncertainties.}
        \label{fig:ue1}
\end{figure*}

\begin{figure*}[!htbp]
    \includegraphics[width=0.27\textwidth]{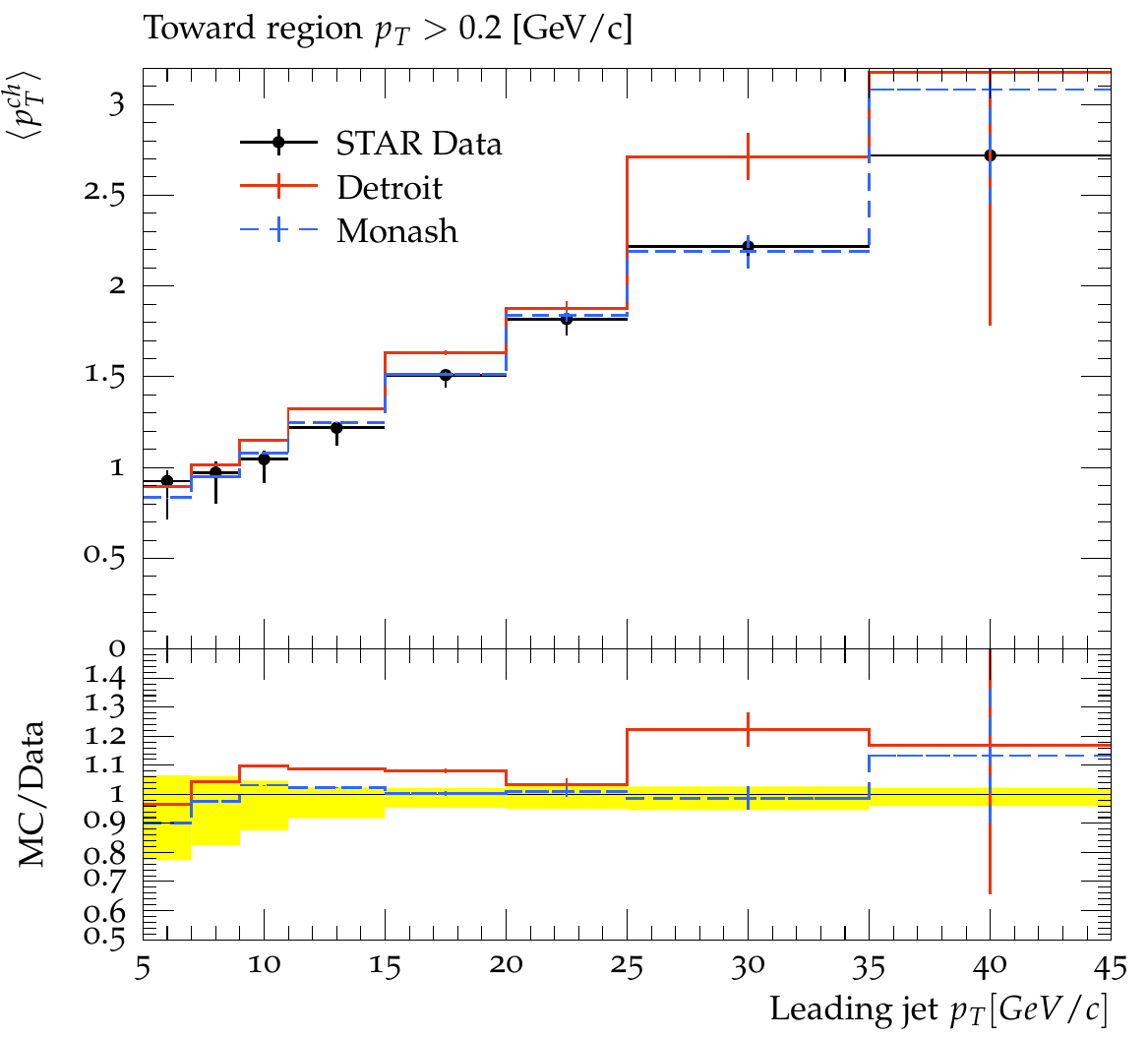}
    \includegraphics[width=0.27\textwidth]{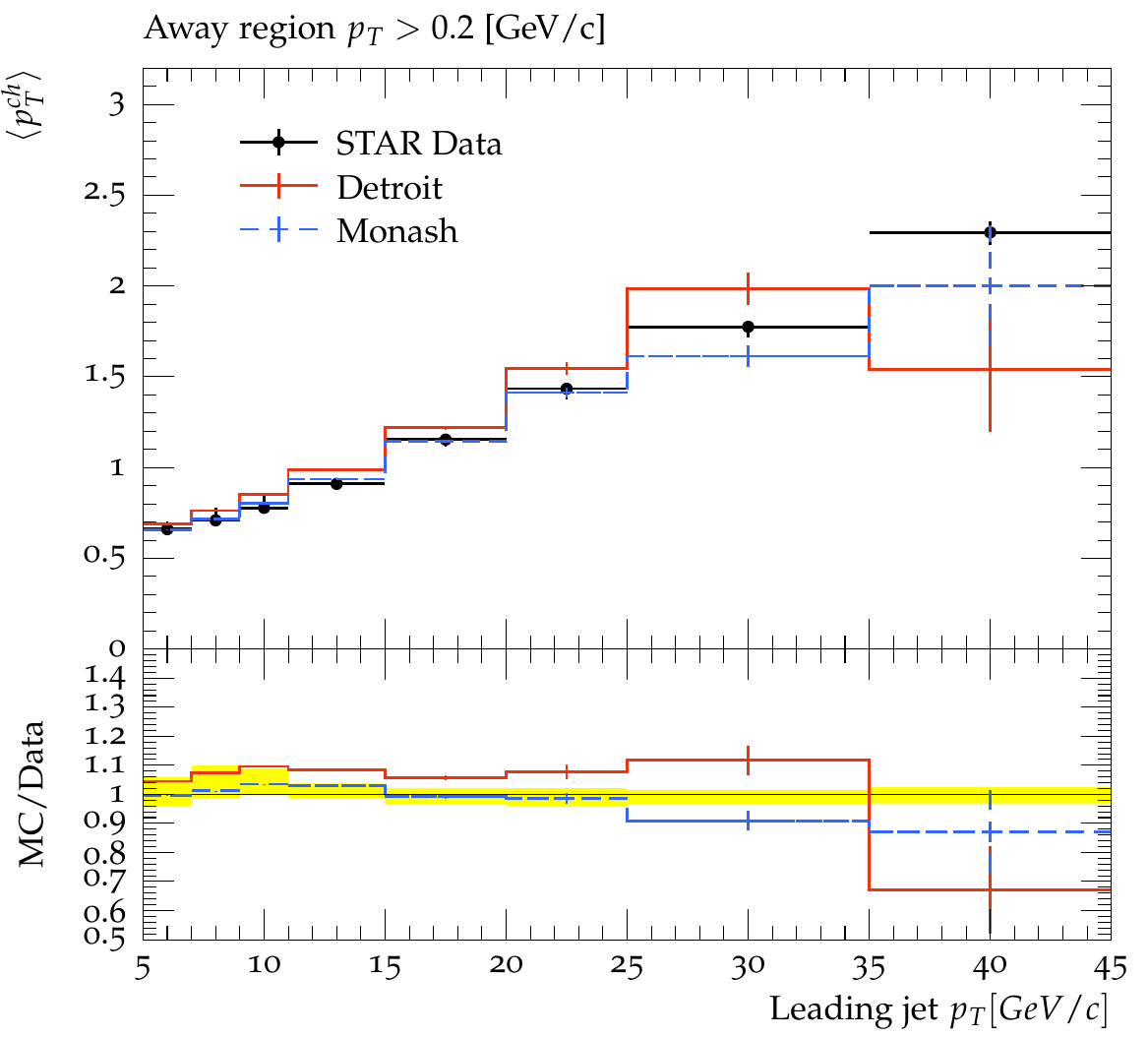}
    \\
    \includegraphics[width=0.27\textwidth]{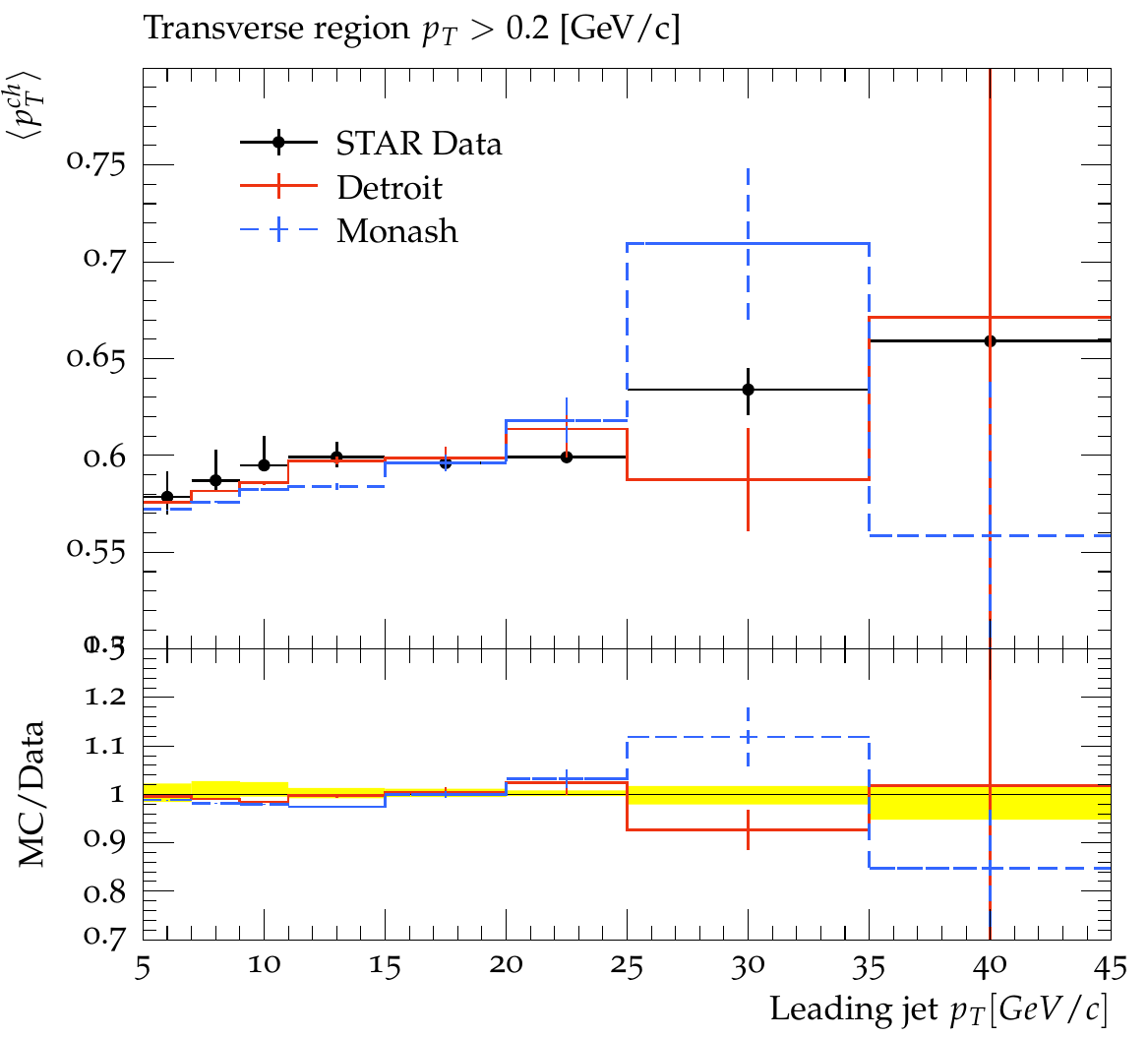}
    \includegraphics[width=0.27\textwidth]{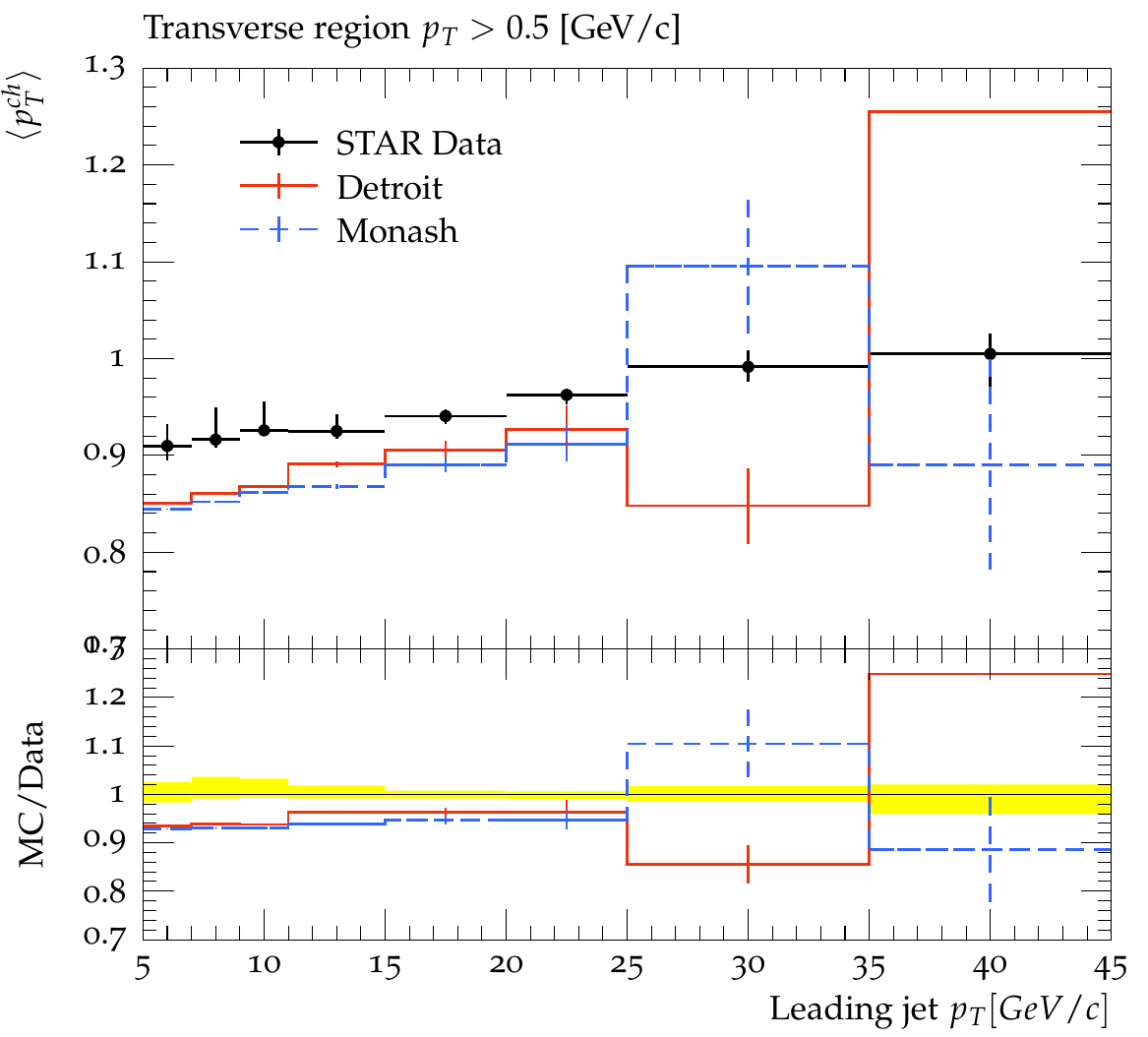}    
    \caption{Average charged particle $p_{T}$ as a function of leading jet in $p$+$p$ collisions at $\s$ = 200 GeV measured by the STAR experiment~\cite{PhysRevD.101.052004}. The top two and bottom left figures show the toward and transverse, and away regions, respectively, for charged particles with $p_{T}>$ 0.2 GeV/$c$. The bottom right shows the transverse region for charged particles with $p_{T}>$ 0.5 GeV/$c$. The bottom panels in each figure show the ratios of the Monte Carlo predictions with respect to the data and the yellow shaded region shows the data uncertainties.}
    \label{fig:ue2}
\end{figure*}

\begin{figure*}[]
    \centering
    \includegraphics[width=0.27\textwidth]{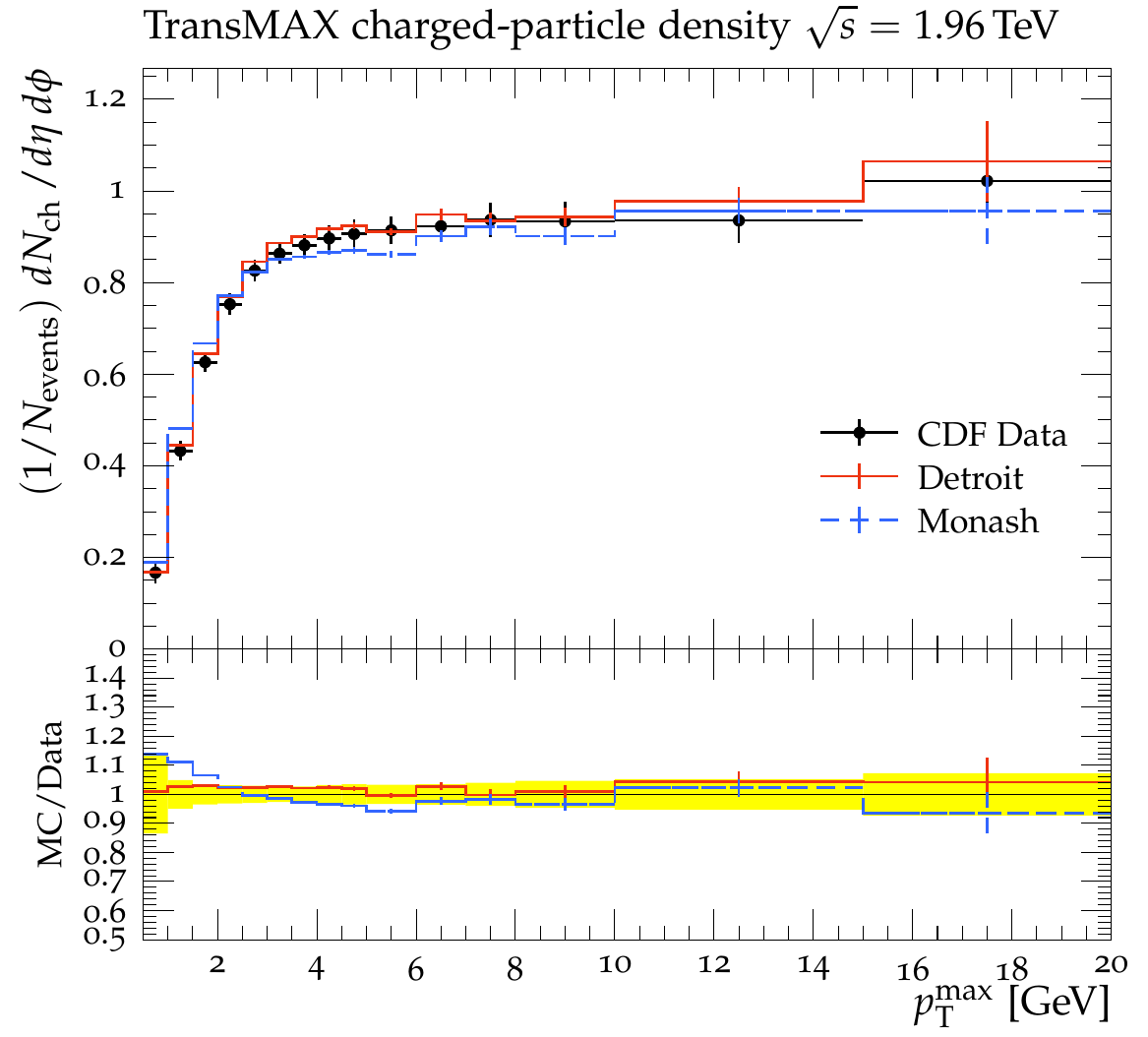}
    \includegraphics[width=0.27\textwidth]{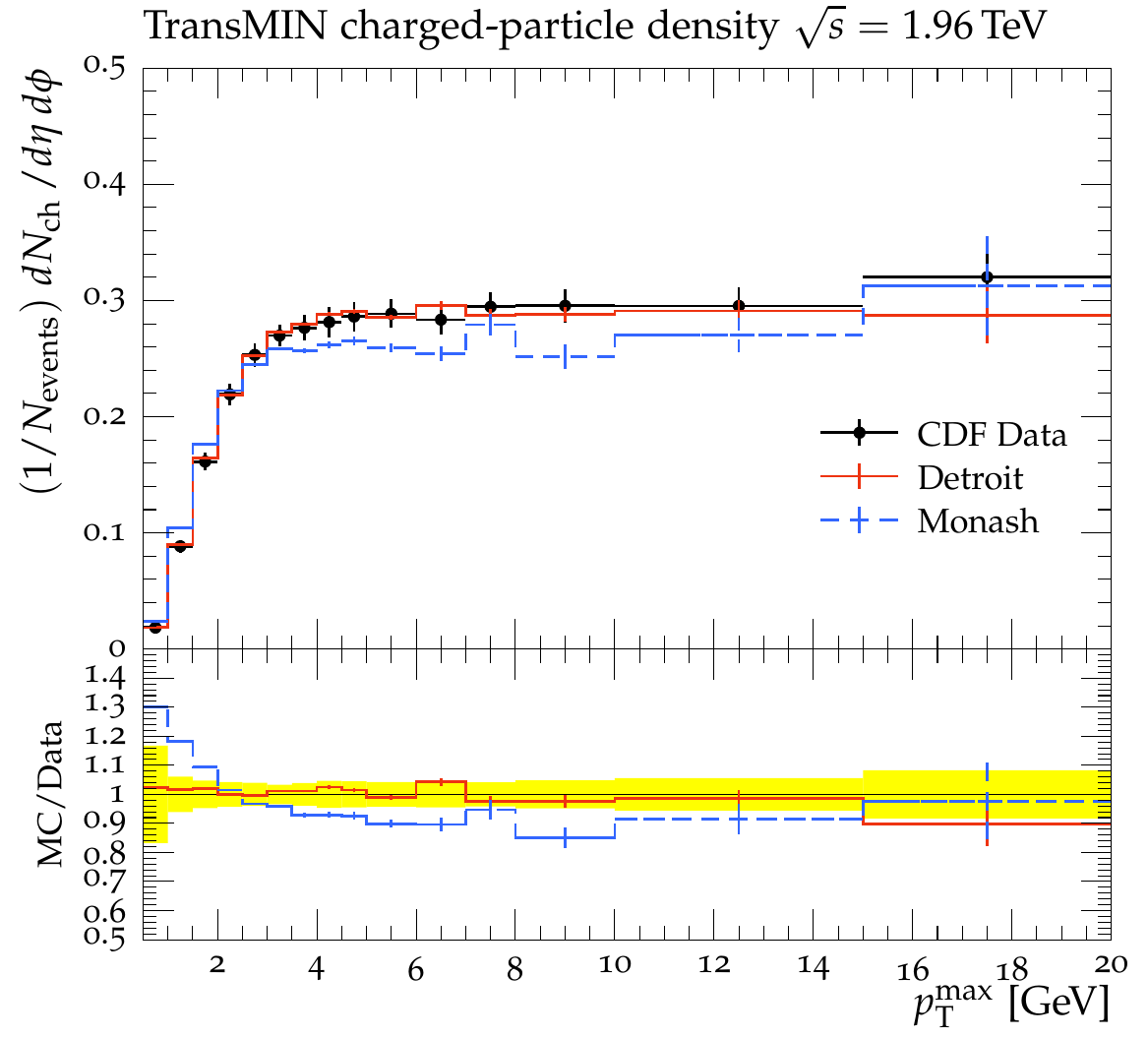}
    \\
    \includegraphics[width=0.27\textwidth]{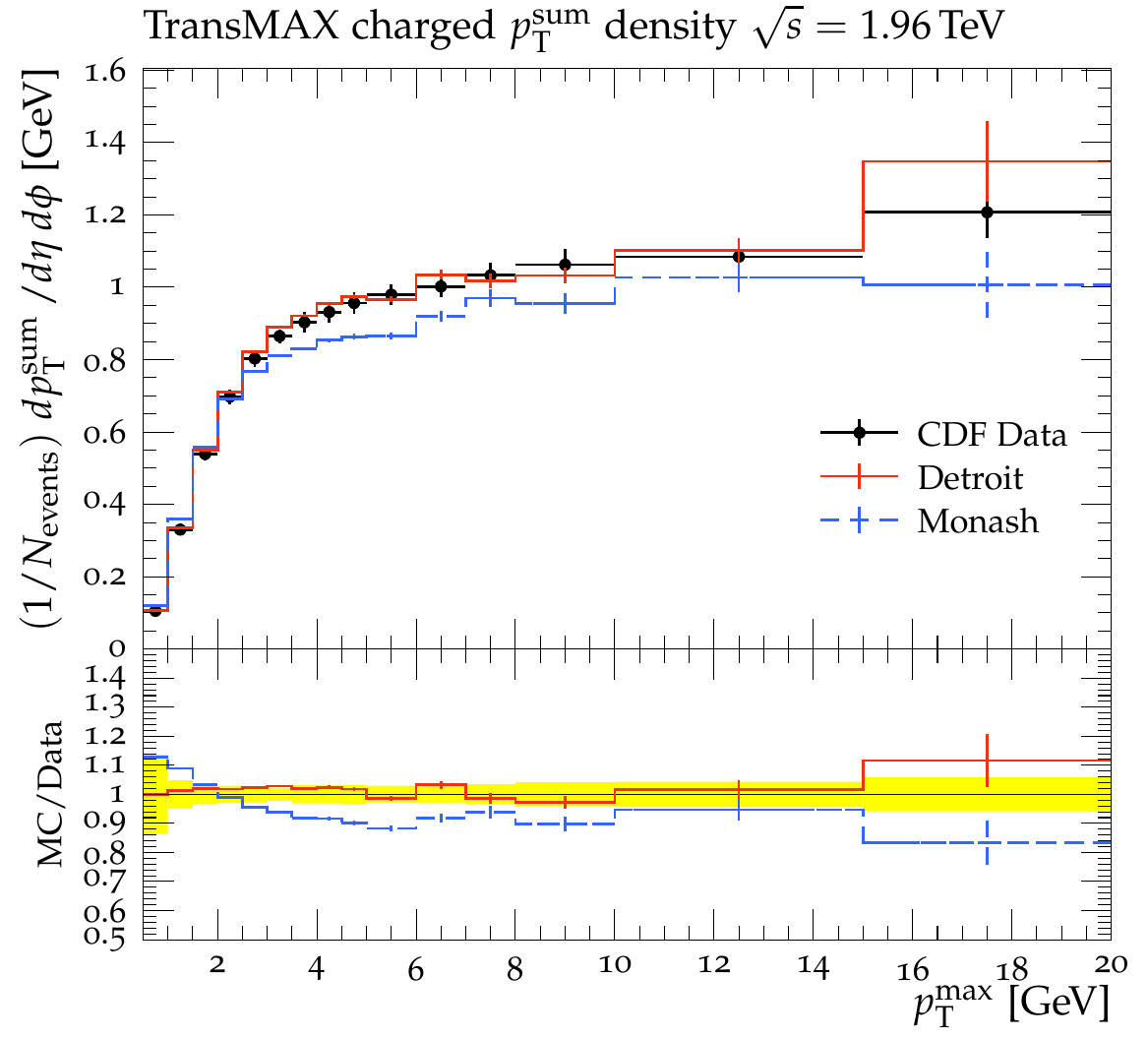}
    \includegraphics[width=0.27\textwidth]{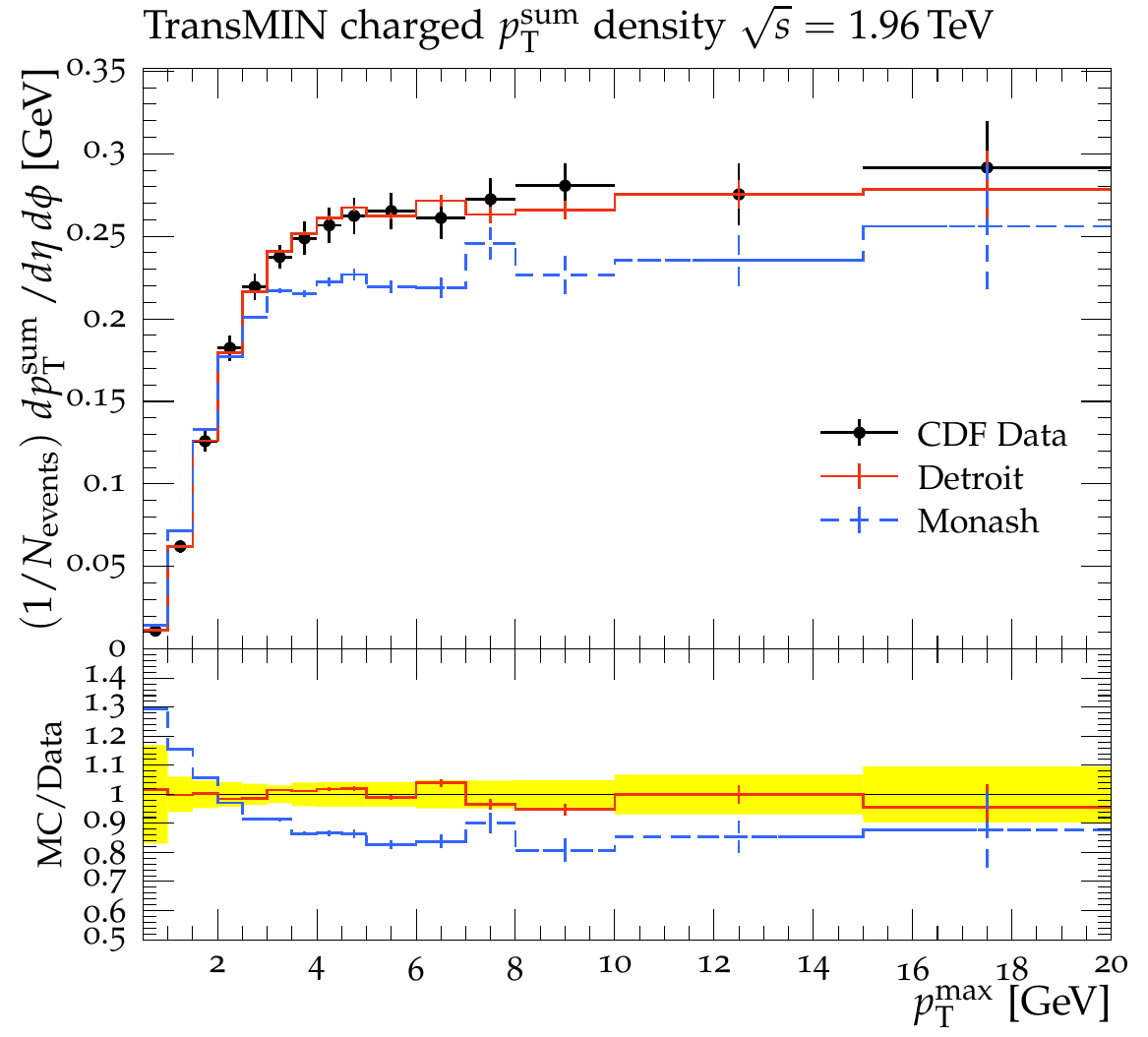}    
    \caption{Underlying event observable as a function of leading hadron $p_{T}$ from the CDF measurement in proton-antiproton collisions at $\s$ = 1960 GeV~\cite{PhysRevD.92.092009}. The top left and right show the charge particle multiplicity in the transMAX and transMIN regions (see text for definitions), respectively. The bottom left and right figures show the charge particle $p_{T}$ sum for the transMAX and transMIN regions, respectively. The bottom panels in each figure show the ratios of the Monte Carlo predictions with respect to the data and the yellow shaded region shows the data uncertainties.}
    \label{fig:cdfue1}
\end{figure*}

\begin{figure*}[]
    \centering
    \includegraphics[width=0.27\textwidth]{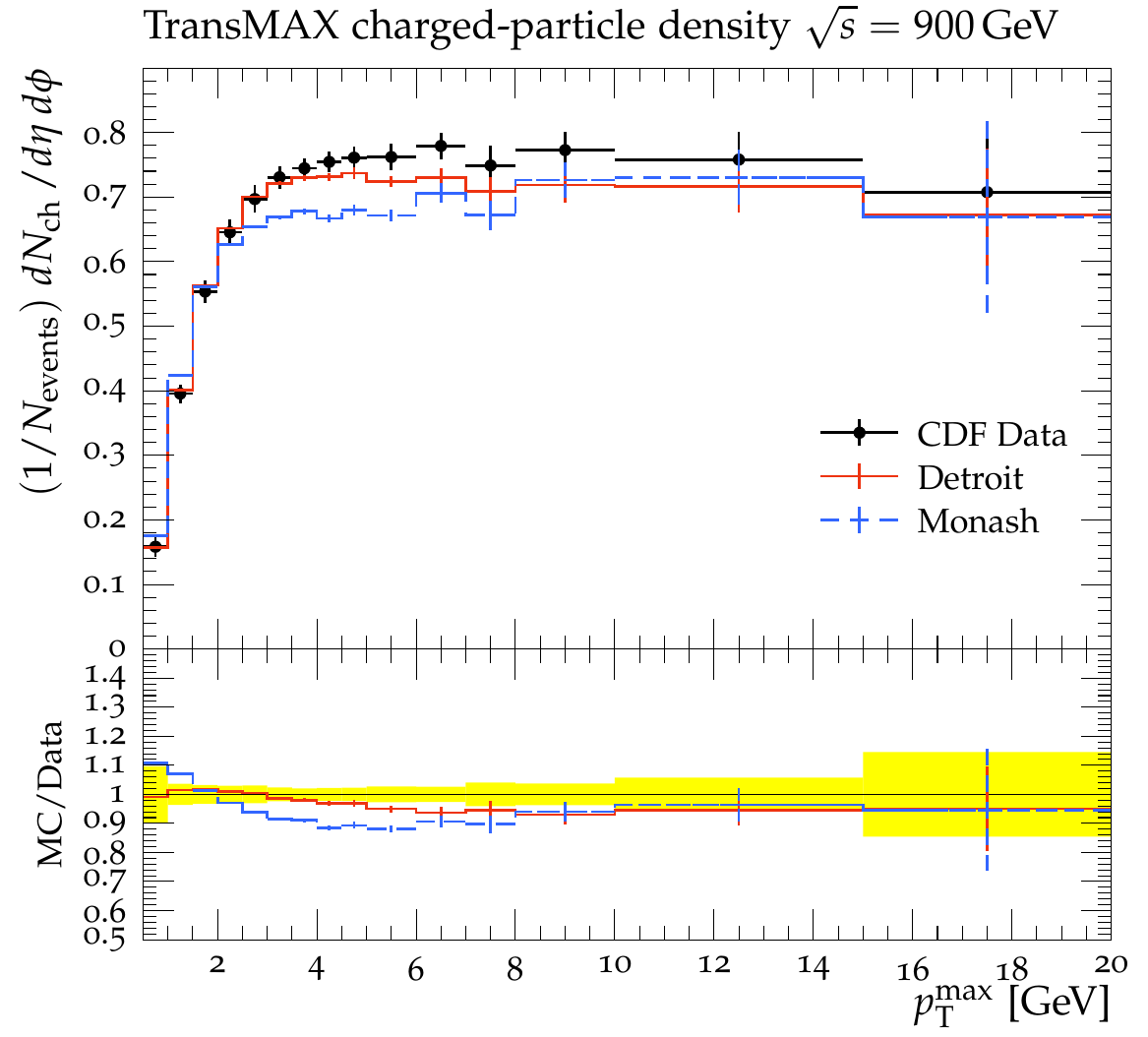}
    \includegraphics[width=0.27\textwidth]{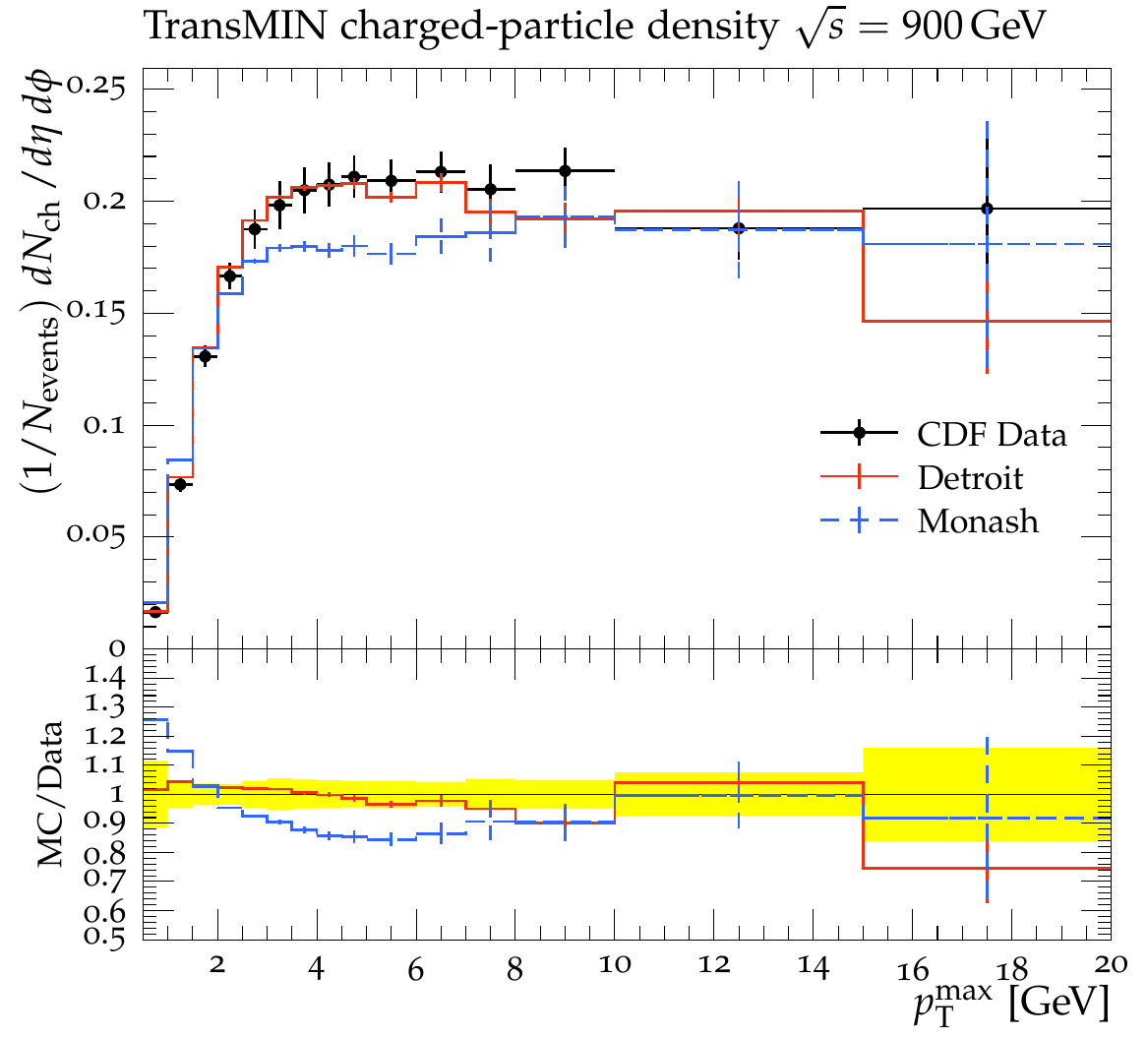}
    \\
    \includegraphics[width=0.27\textwidth]{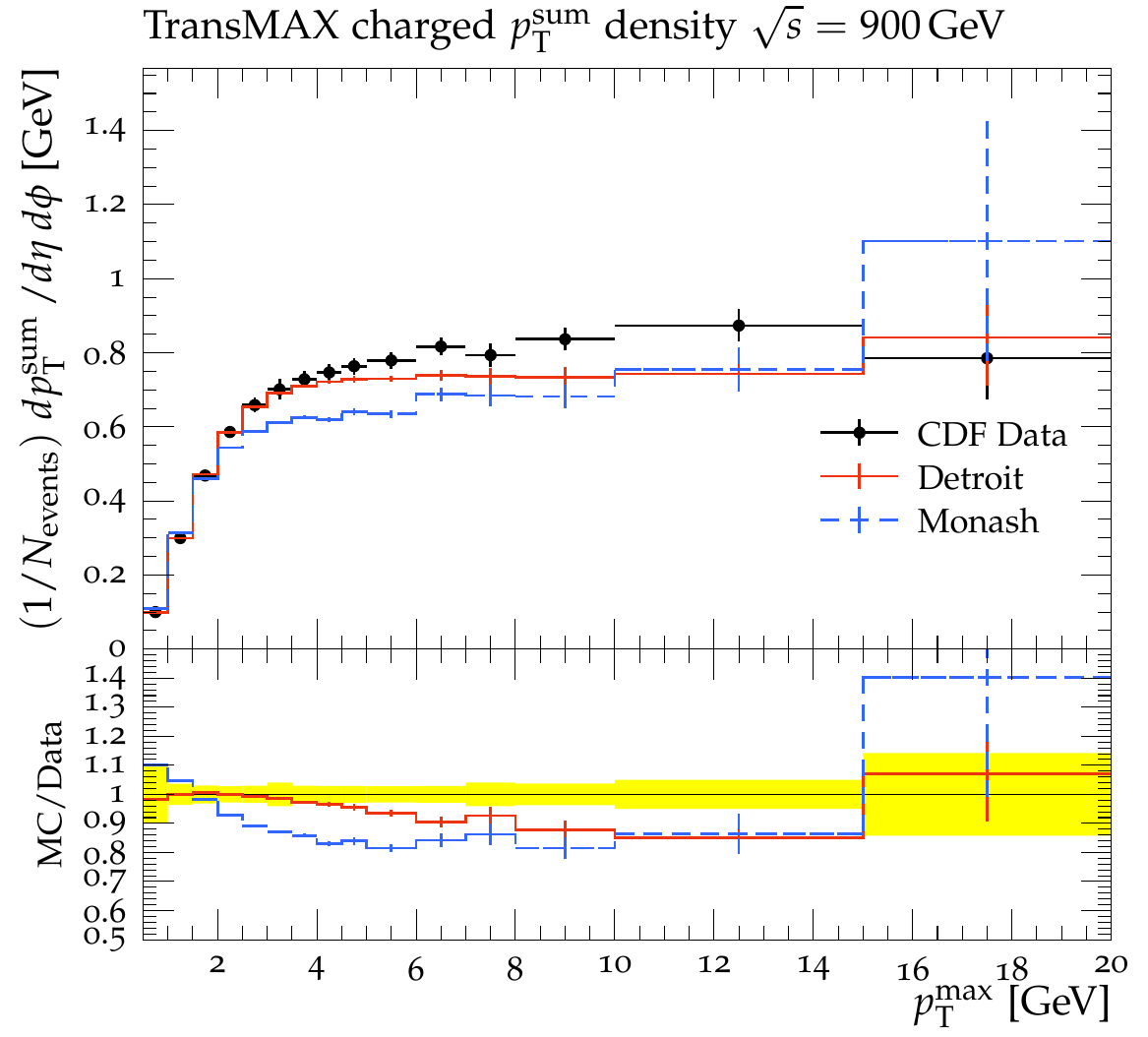}
    \includegraphics[width=0.27\textwidth]{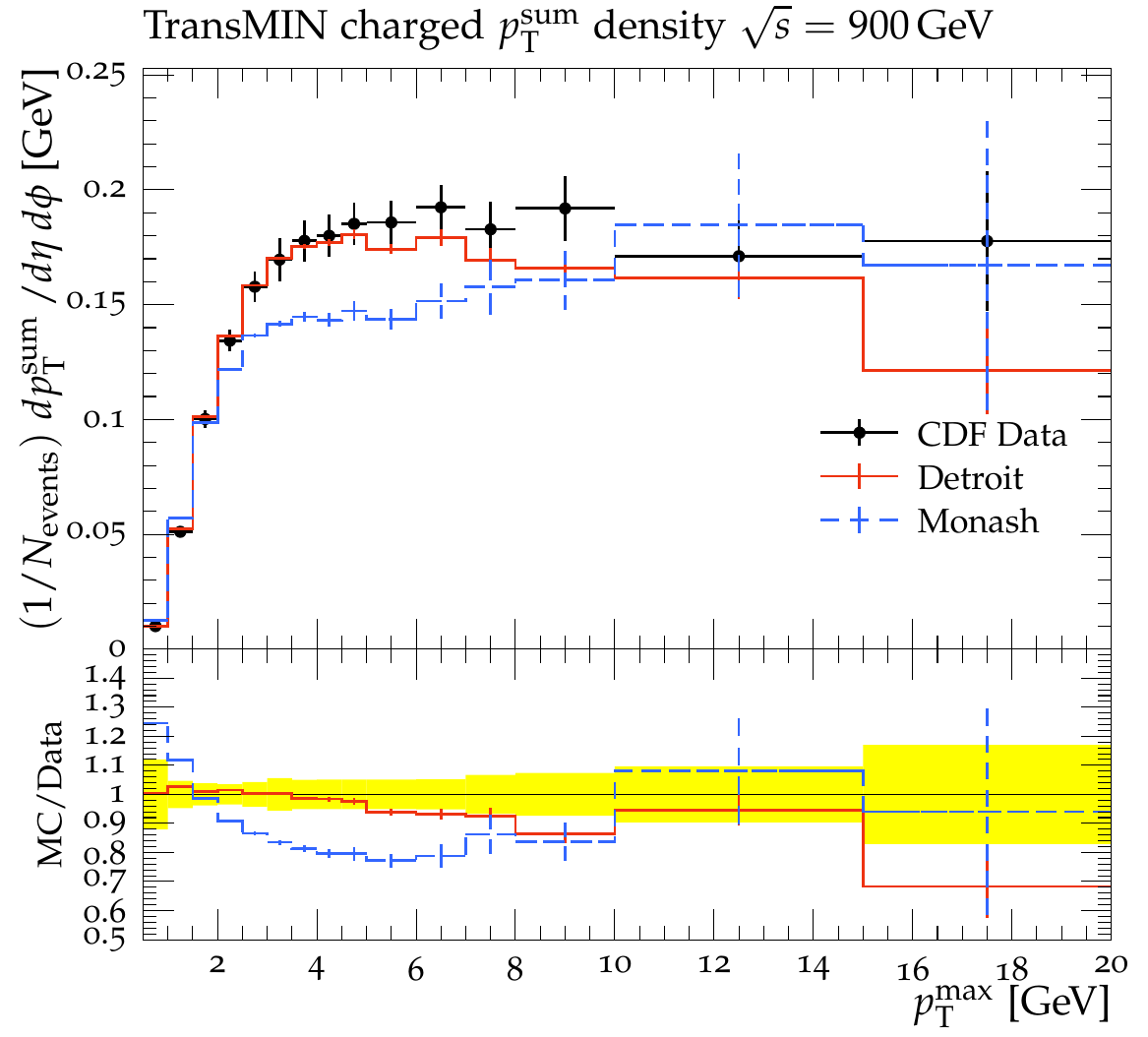}    
    \caption{Underlying event observable as a function of leading hadron $p_{T}$ from the CDF measurement in proton-antiproton collisions at $\s$ = 900 GeV~\cite{PhysRevD.92.092009}. The top left and right show the charge particle multiplicity in the transMAX and transMIN regions (see text for definitions), respectively. The bottom left and right figures show the charge particle $p_{T}$ sum for the transMAX and transMIN regions, respectively. The bottom panels in each figure show the ratios of the Monte Carlo predictions with respect to the data and the yellow shaded region shows the data uncertainties.}
    \label{fig:cdfue2}
\end{figure*}
\begin{figure*}[!htbp]
    \centering
    \includegraphics[width=0.27\textwidth]{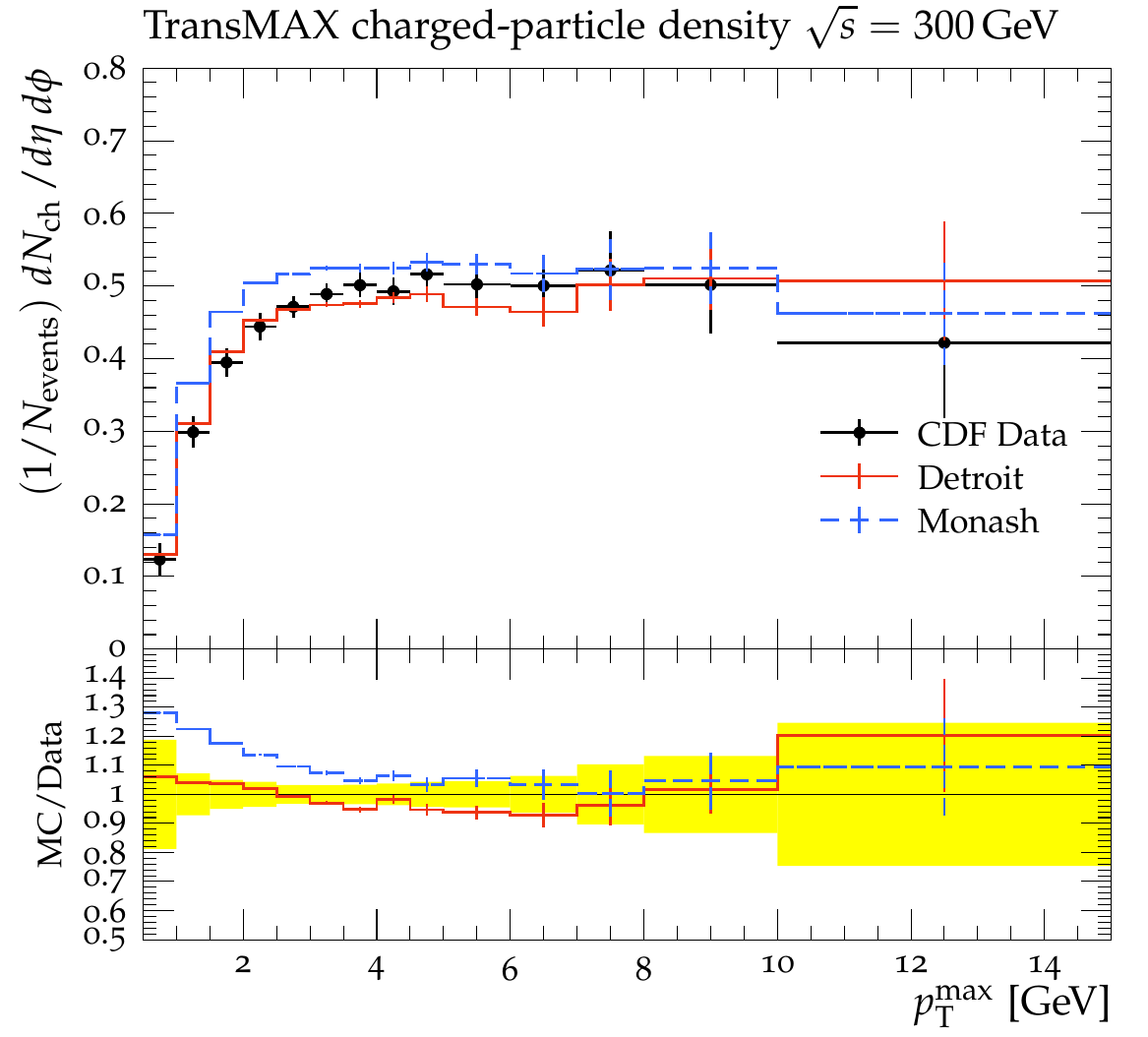}
    \includegraphics[width=0.27\textwidth]{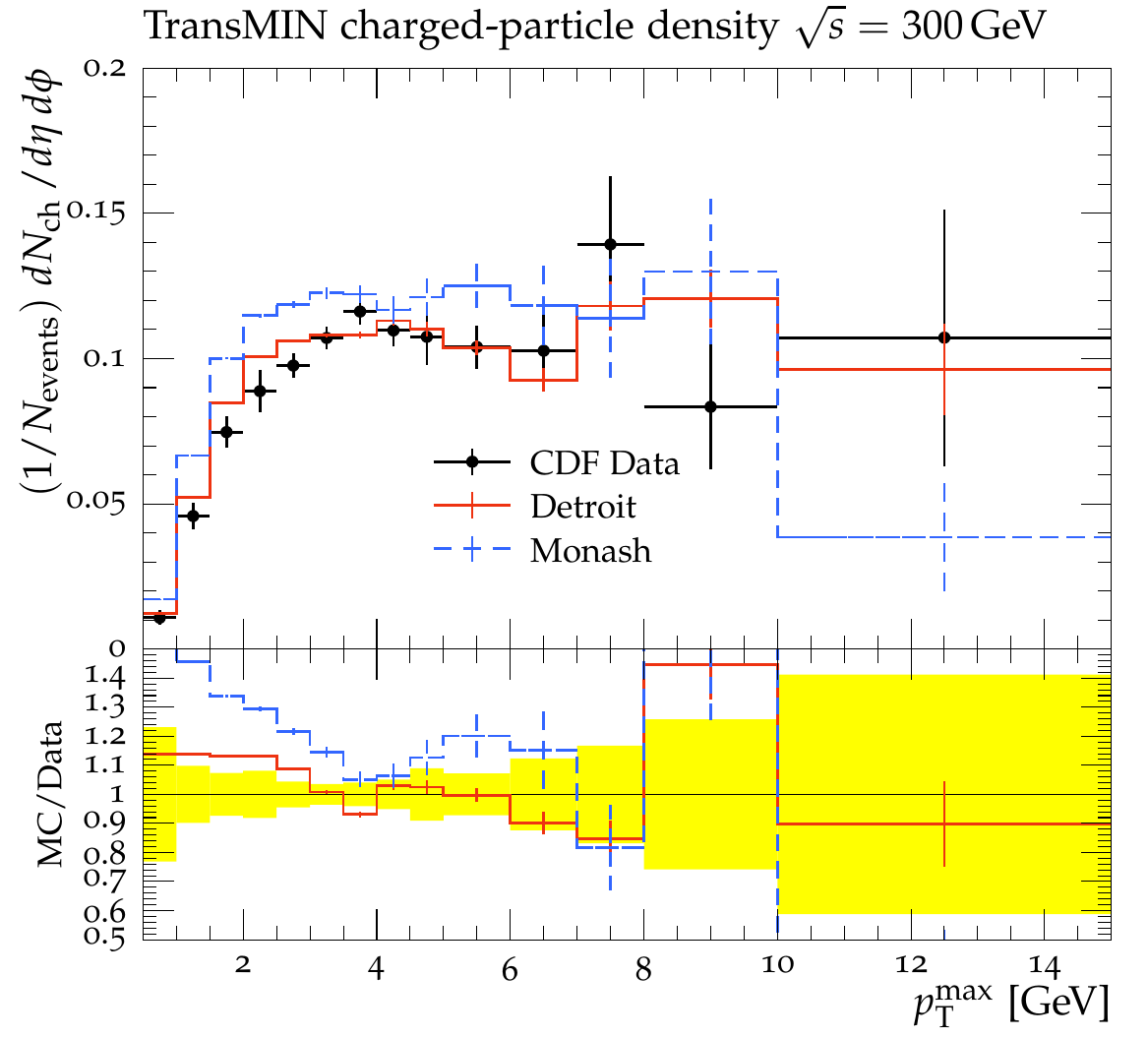}
    \\
    \includegraphics[width=0.27\textwidth]{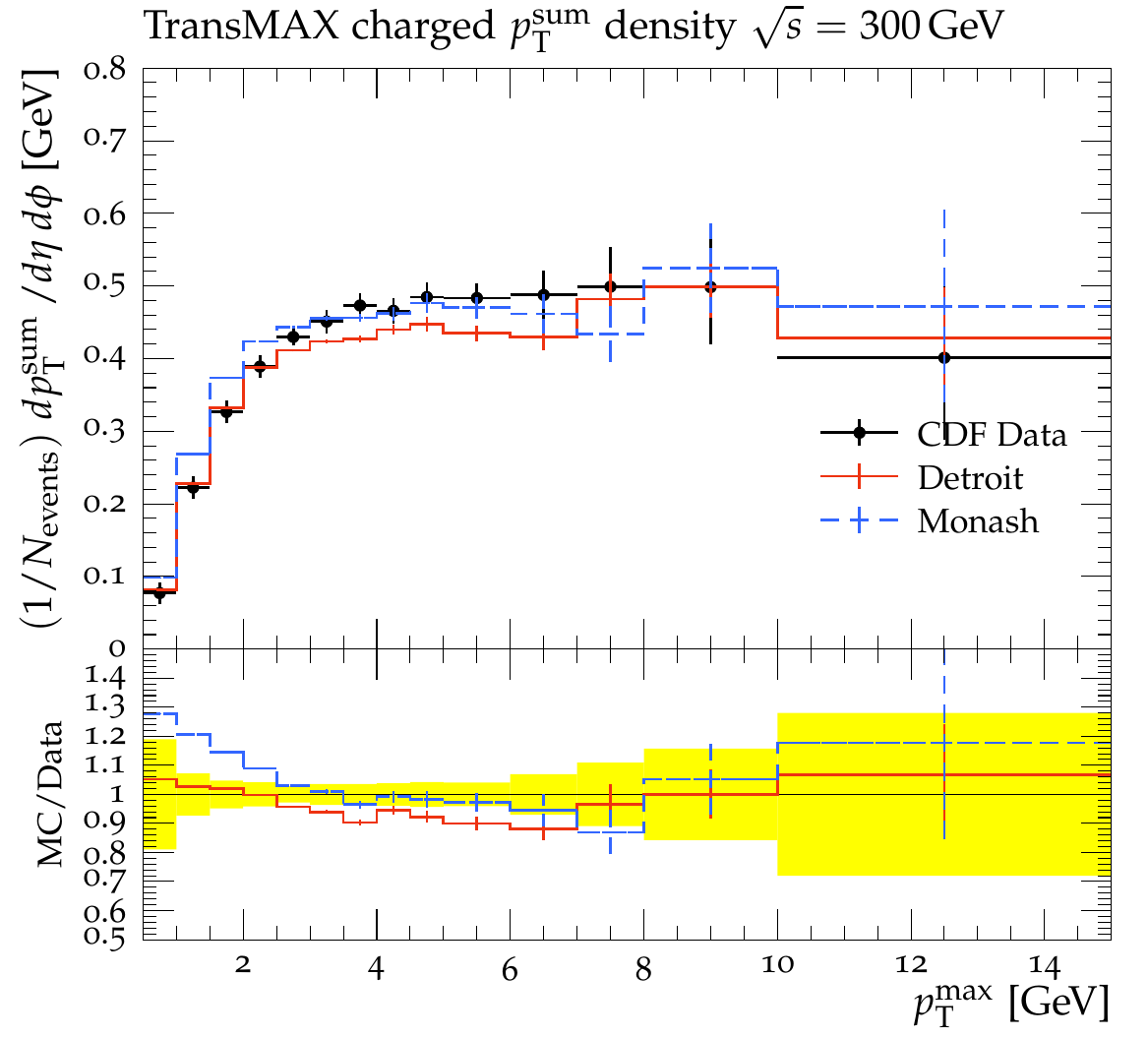}
    \includegraphics[width=0.27\textwidth]{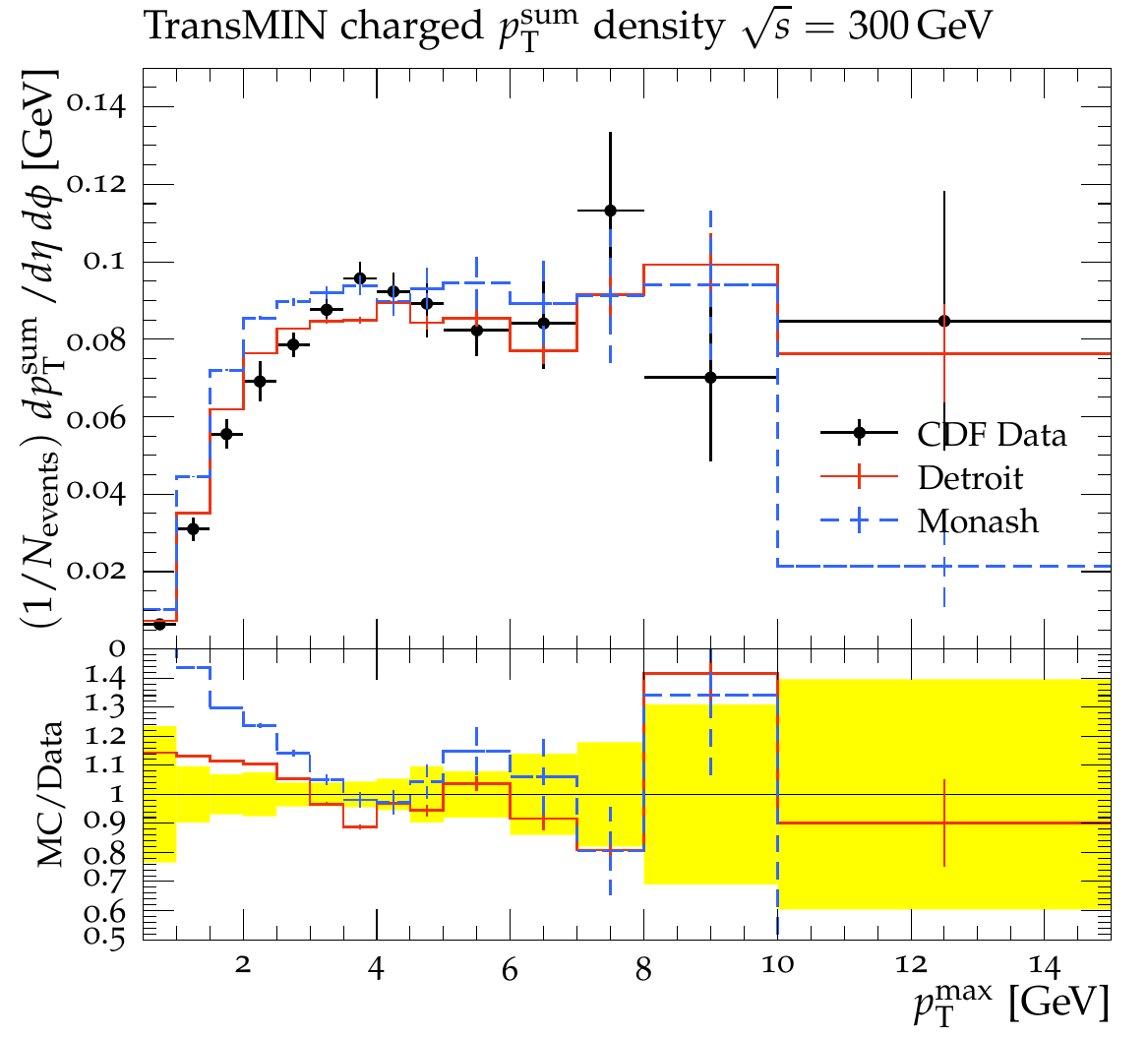}    
    \caption{Underlying event observable as a function of leading hadron $p_{T}$ from the CDF measurement in proton-antiproton collisions at $\s$ = 300 GeV~\cite{PhysRevD.92.092009}. The top two figures show the charge particle multiplicity in the transMAX and transMIN regions (see text for definitions). The bottom two figures show the charge particle $p_{T}$ sum for the transMAX and transMIN regions. The bottom panels in each figure show the ratios of the Monte Carlo predictions with respect to the data and the yellow shaded region shows the data uncertainties.}
    \label{fig:cdfue3}
\end{figure*}

\begin{figure*}[!htbp]
    \centering
    \includegraphics[width=0.27\textwidth]{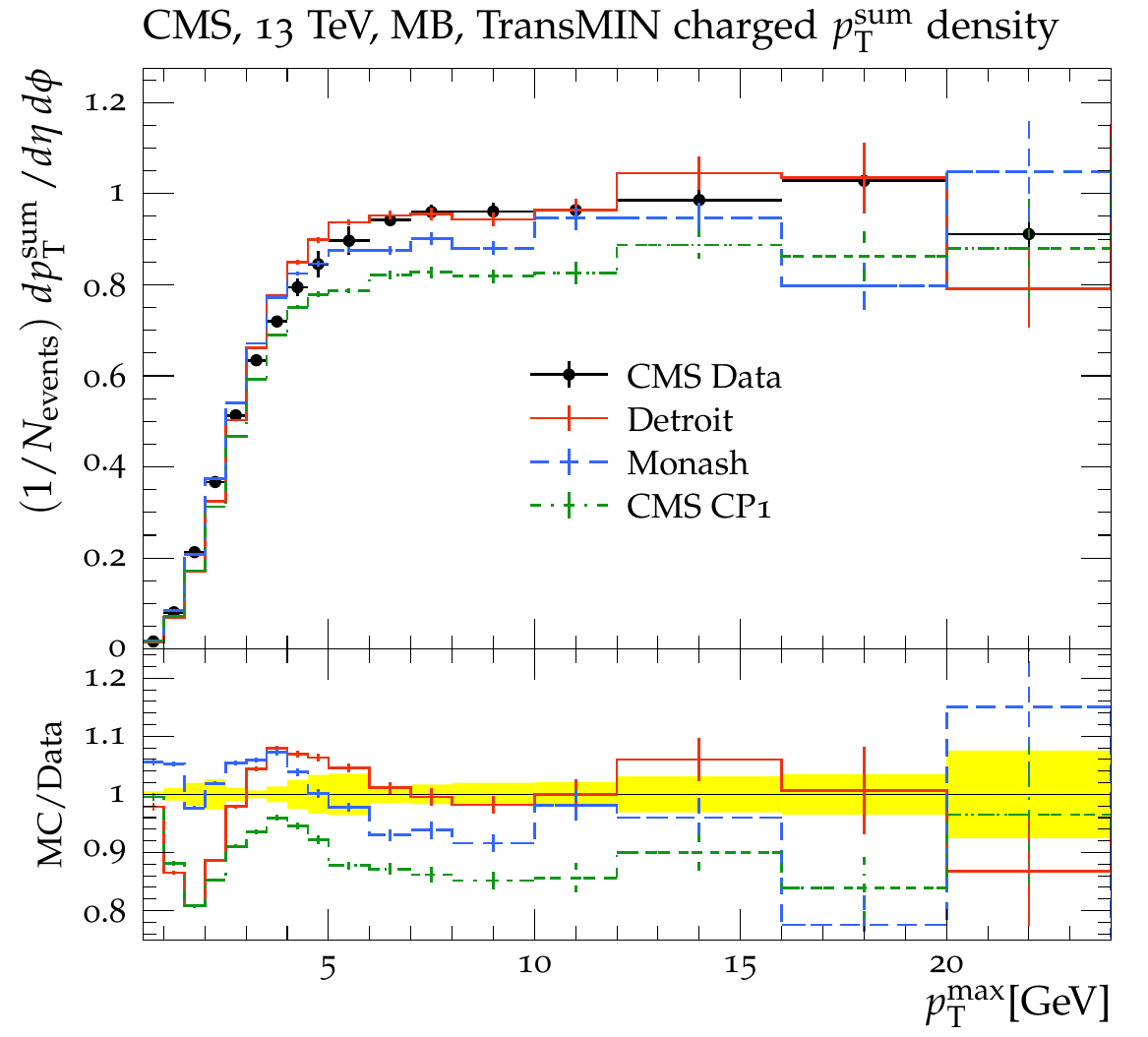}
    \includegraphics[width=0.27\textwidth]{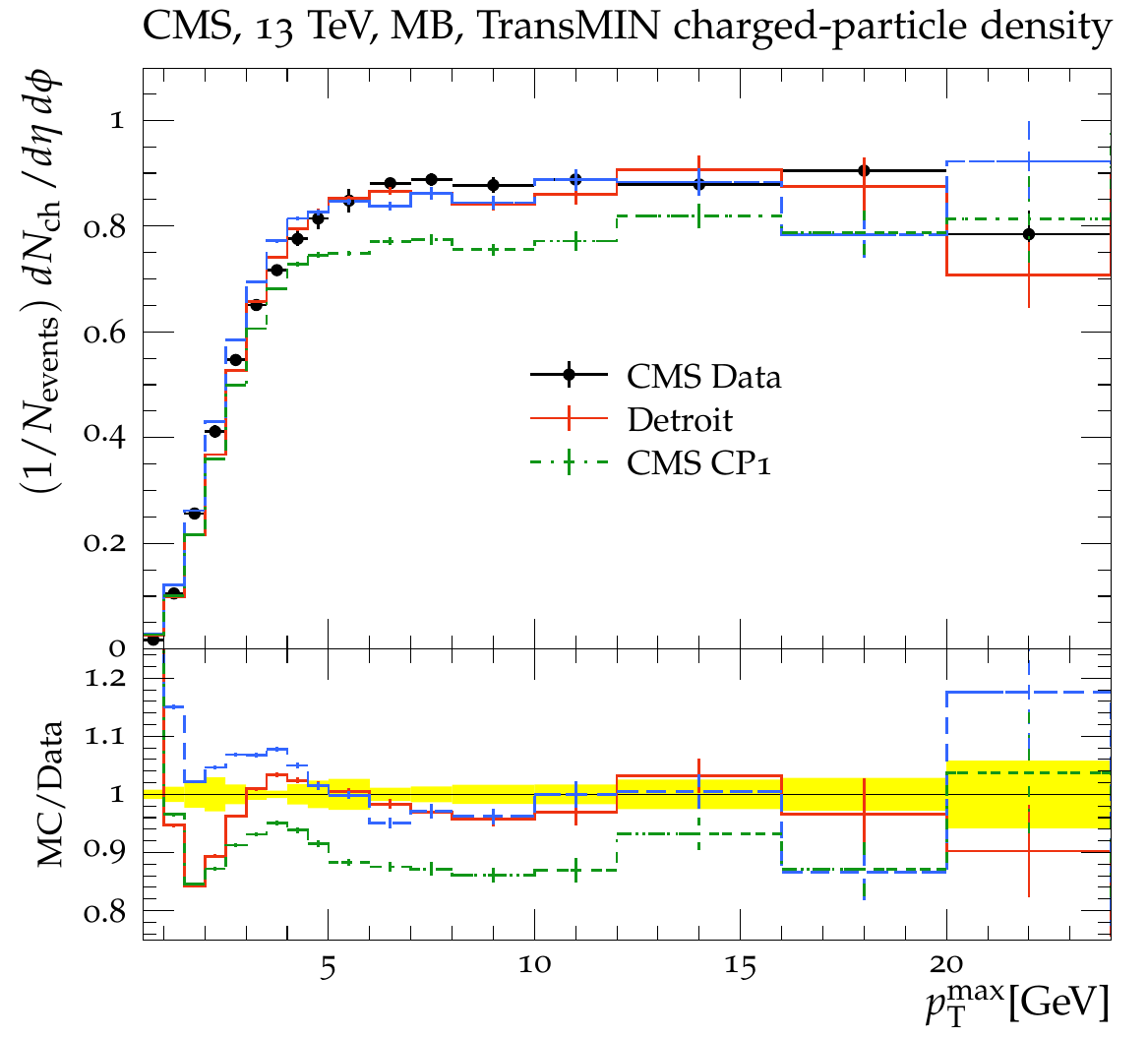}
    \\
    \includegraphics[width=0.27\textwidth]{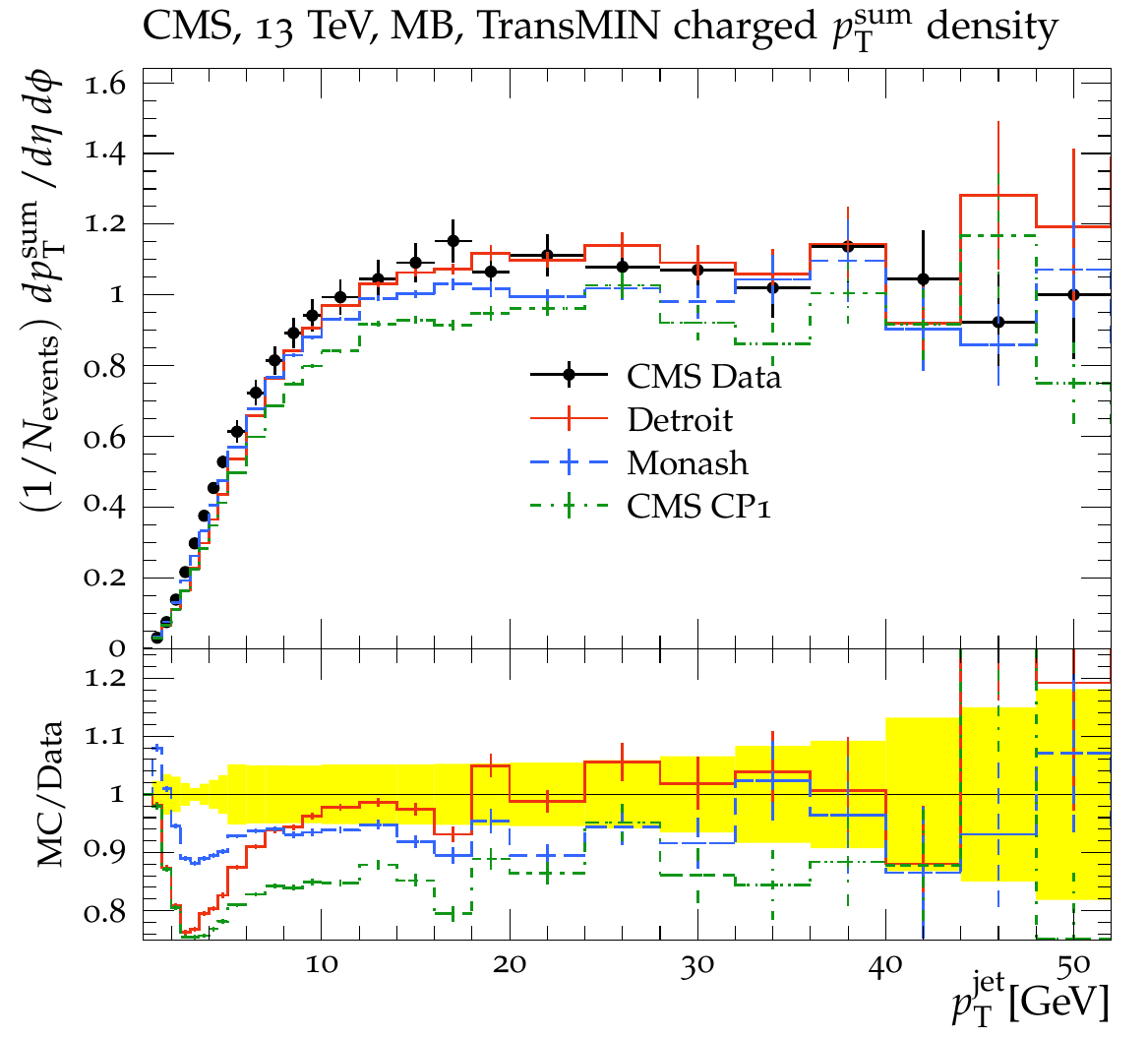}
    \includegraphics[width=0.27\textwidth]{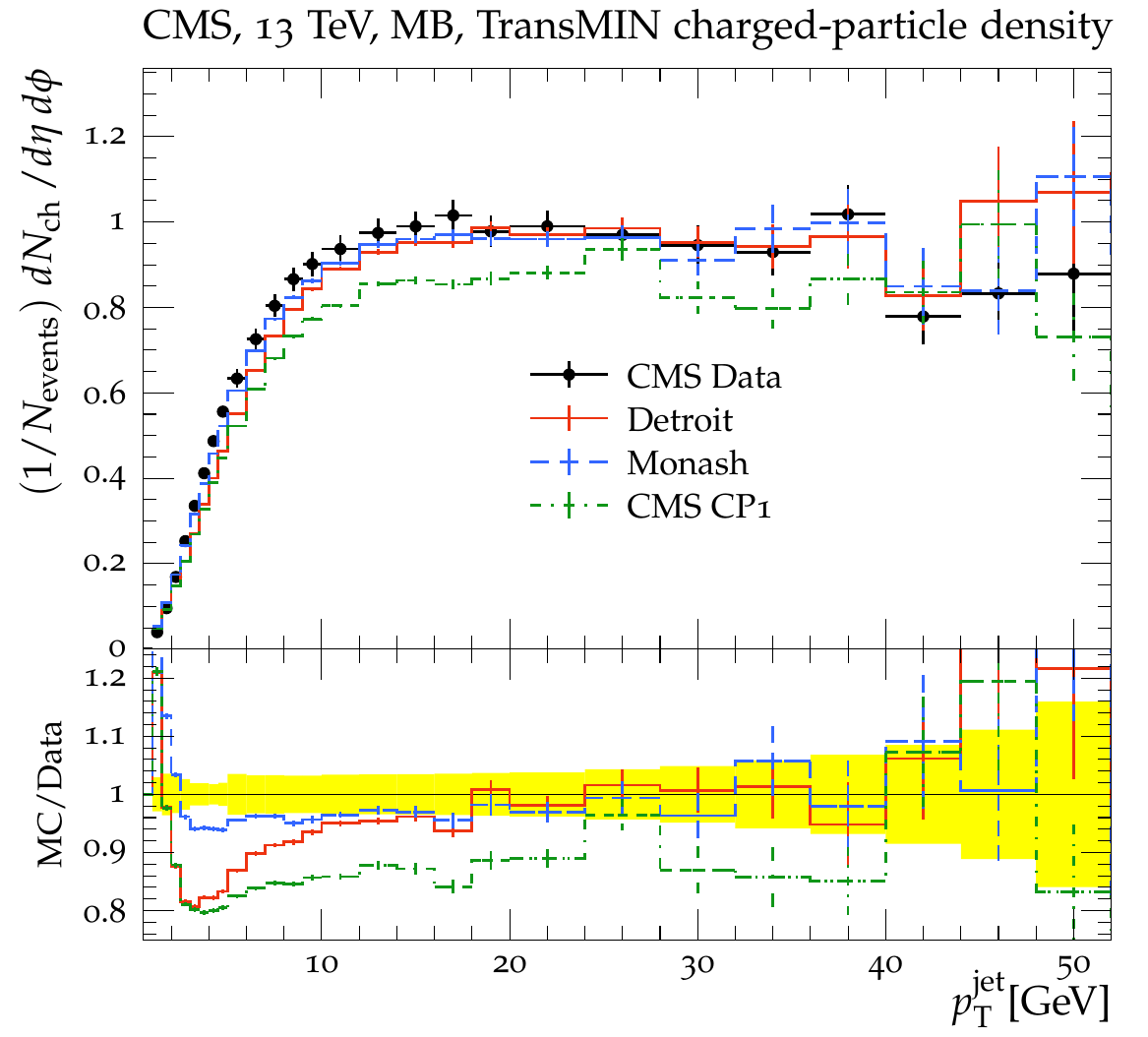}
    \caption{The $p_{T}$ sum (left) and charge particle multiplicity (right) as a function of leading track (top) and jet(bottom) $p_{T}$ in the transMin region from the CMS measurement in proton-proton collisions at $\s$ = 13 TeV~\cite{CMS:2015zev}. The bottom panels in each figure show the ratios of the Monte Carlo predictions with respect to the data and the yellow shaded region shows the data uncertainties.}
    \label{fig:cms3}
\end{figure*}
\begin{figure}[!htbp]
    \centering
    \includegraphics[width=0.35\textwidth]{figs/STAR_2006_PID_d02-x01-y01.pdf}
    \includegraphics[width=0.35\textwidth]{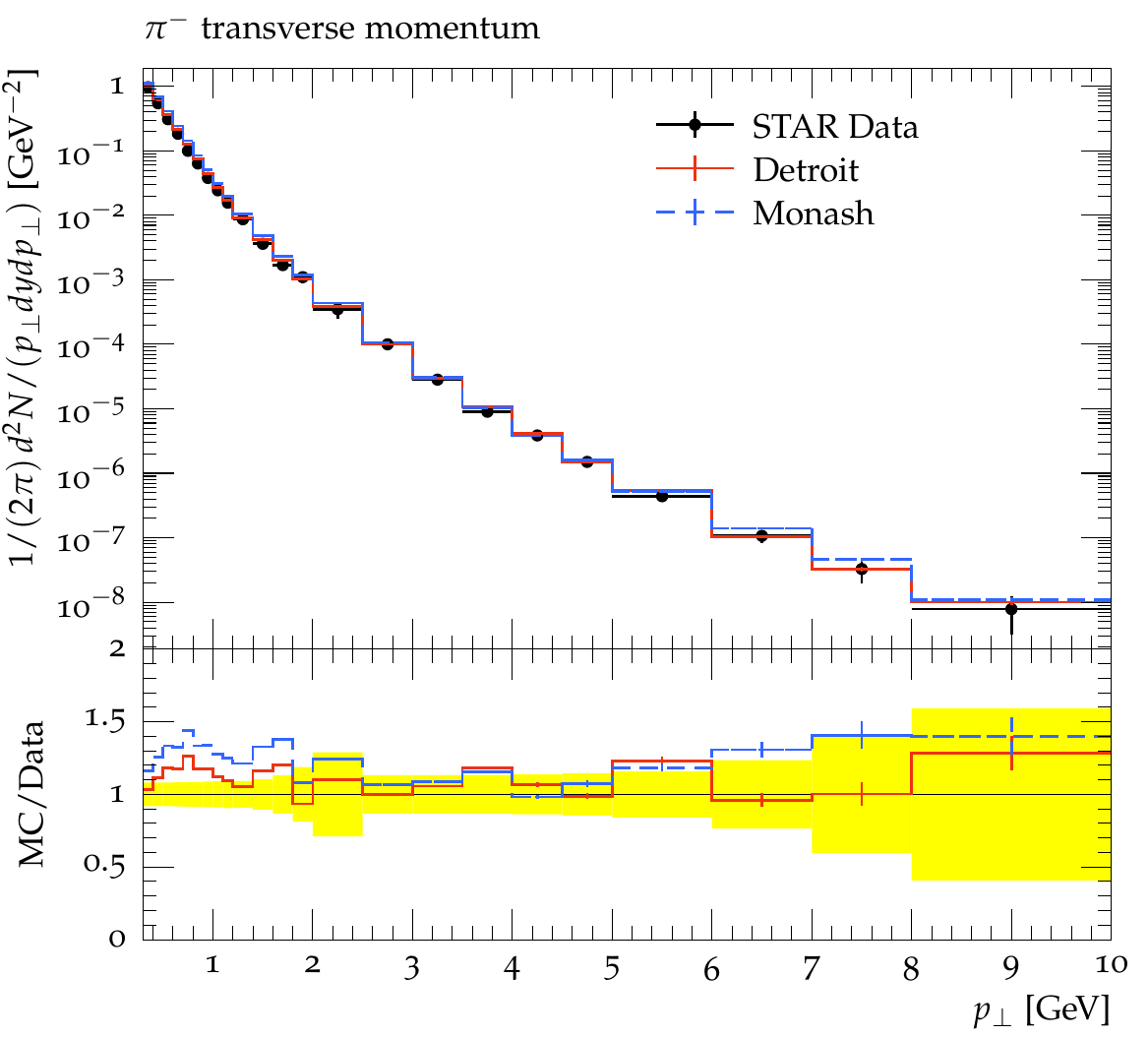}
    \caption{Mid-rapidity $\pi$ cross sections as a function of $p_{T}$ in $p$+$p$ collisions at $\s$ = 200 GeV~\cite{ADAMS2006161}. The left figure shows $\pi^{+}$ and the right $\pi^{-}$. The bottom panels in each figure show the ratios of the Monte Carlo predictions with respect to the data and the yellow shaded region shows the data uncertainties.}
    \label{fig:pid}
\end{figure}
\begin{figure*}[!htbp]
    \centering
    \includegraphics[width=0.35\textwidth]{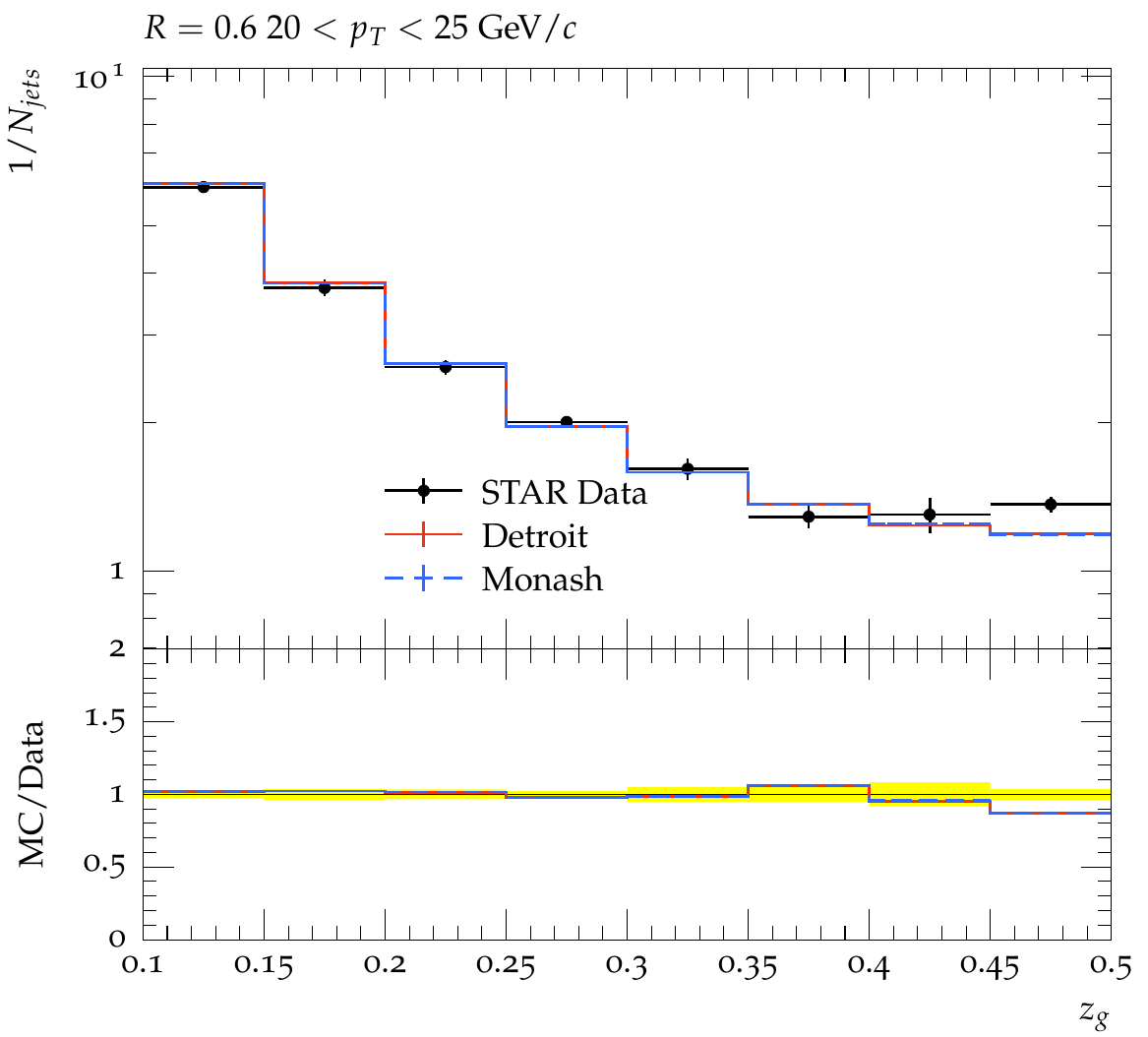}
    \includegraphics[width=0.35\textwidth]{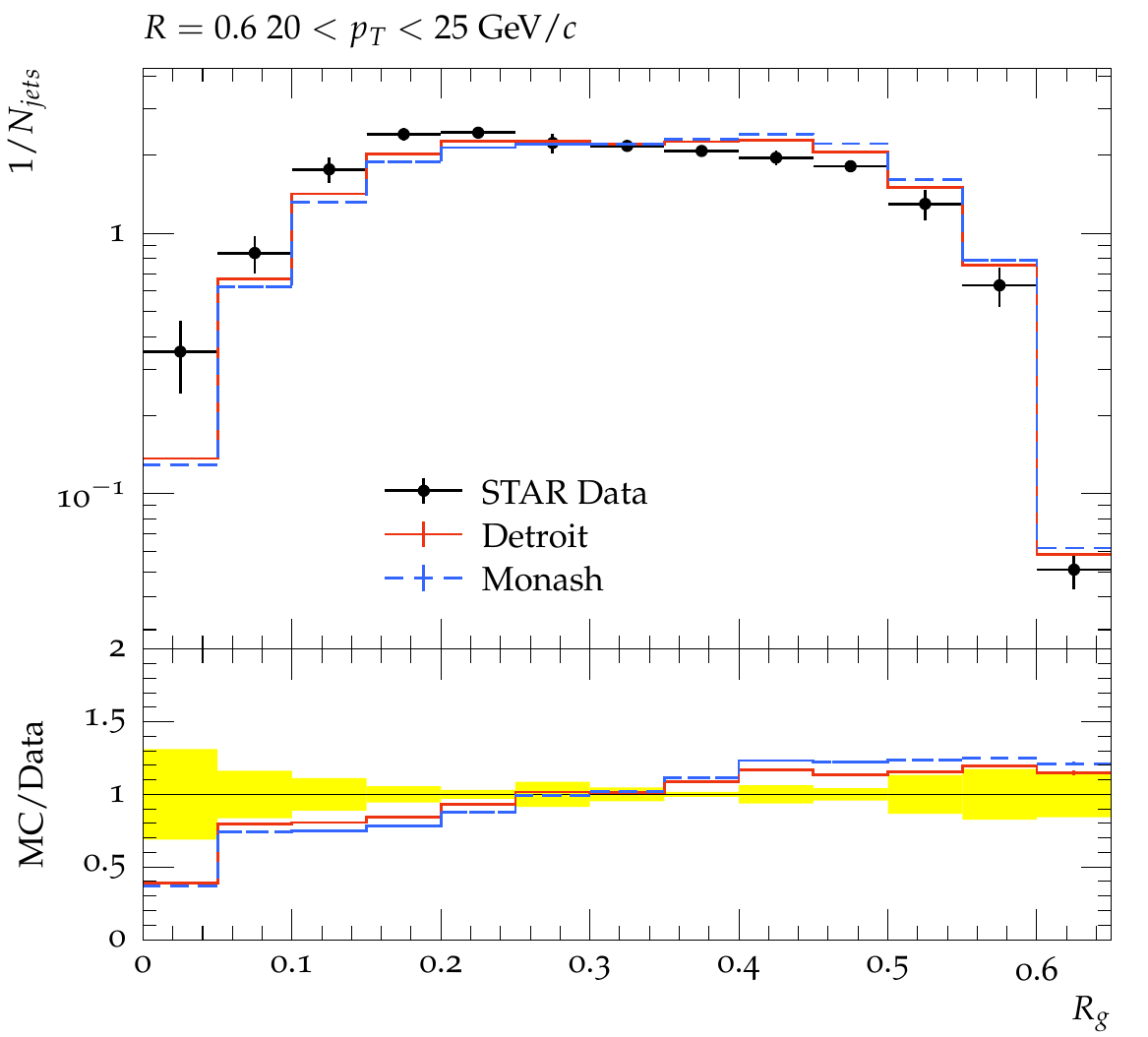} 
    \includegraphics[width=0.35\textwidth]{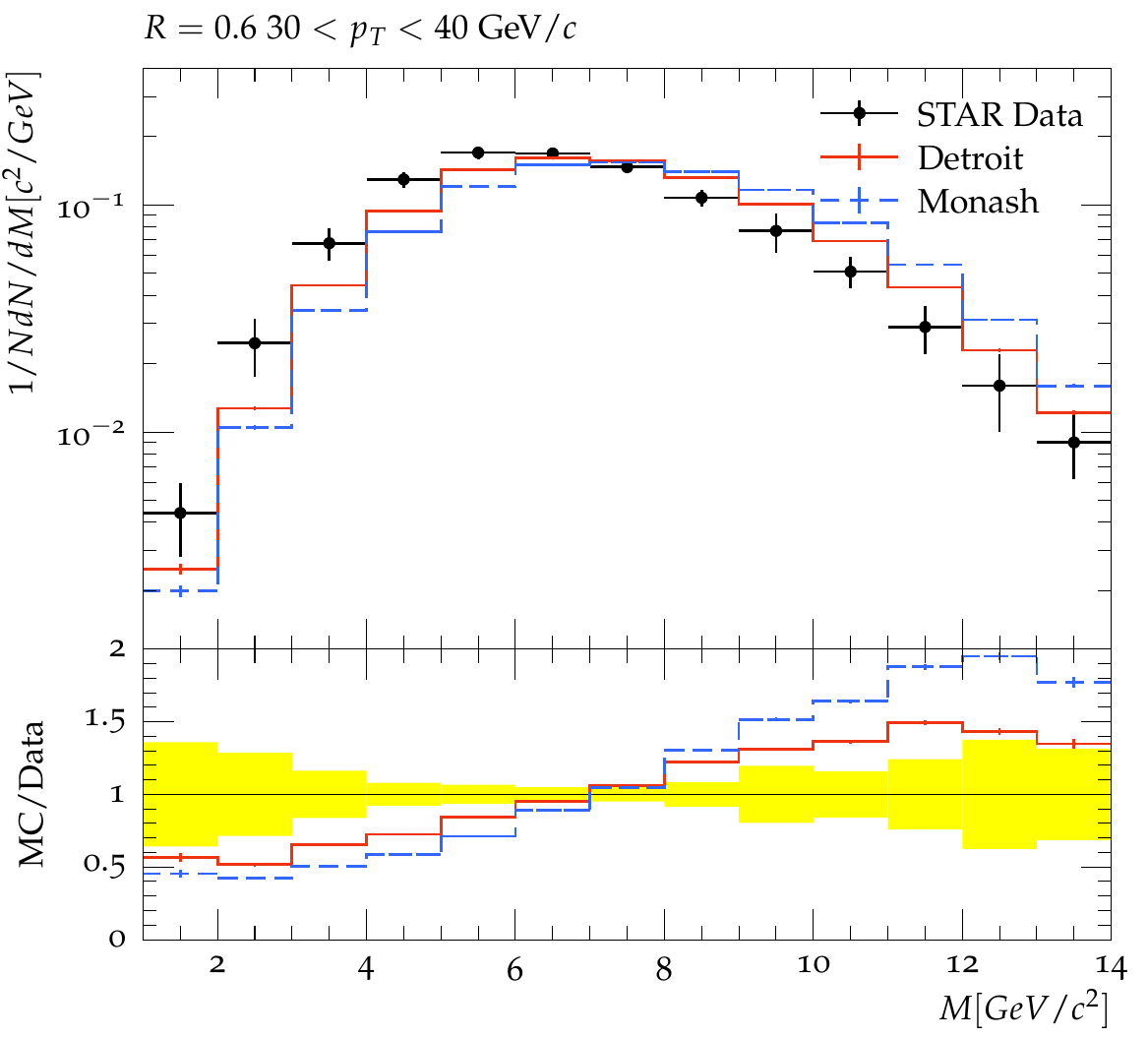}
    \caption{Example jet sub-structure distributions in $p$+$p$ collisions at $\s$ = 200 GeV measured by the STAR experiment. The top left and right figures show the $z_{g}$ and $R_{g}$ observables for anti-kT jets with R=0.6 and jet $p_{T}$ 20 to 25 GeV/$c$~\cite{ADAM2020135846}. The bottom shows the jet mass for anti-kT jets with R=0.6 and $p_{T}$ 30 to 40 GeV/$c$~\cite{starcollaboration2021invariant}. The bottom panels in each figure show the ratios of the Monte Carlo predictions with respect to the data and the yellow shaded region shows the data uncertainties.}
    \label{fig:jet}
\end{figure*}

\begin{figure*}[]
    \centering
    \includegraphics[width=0.35\textwidth]{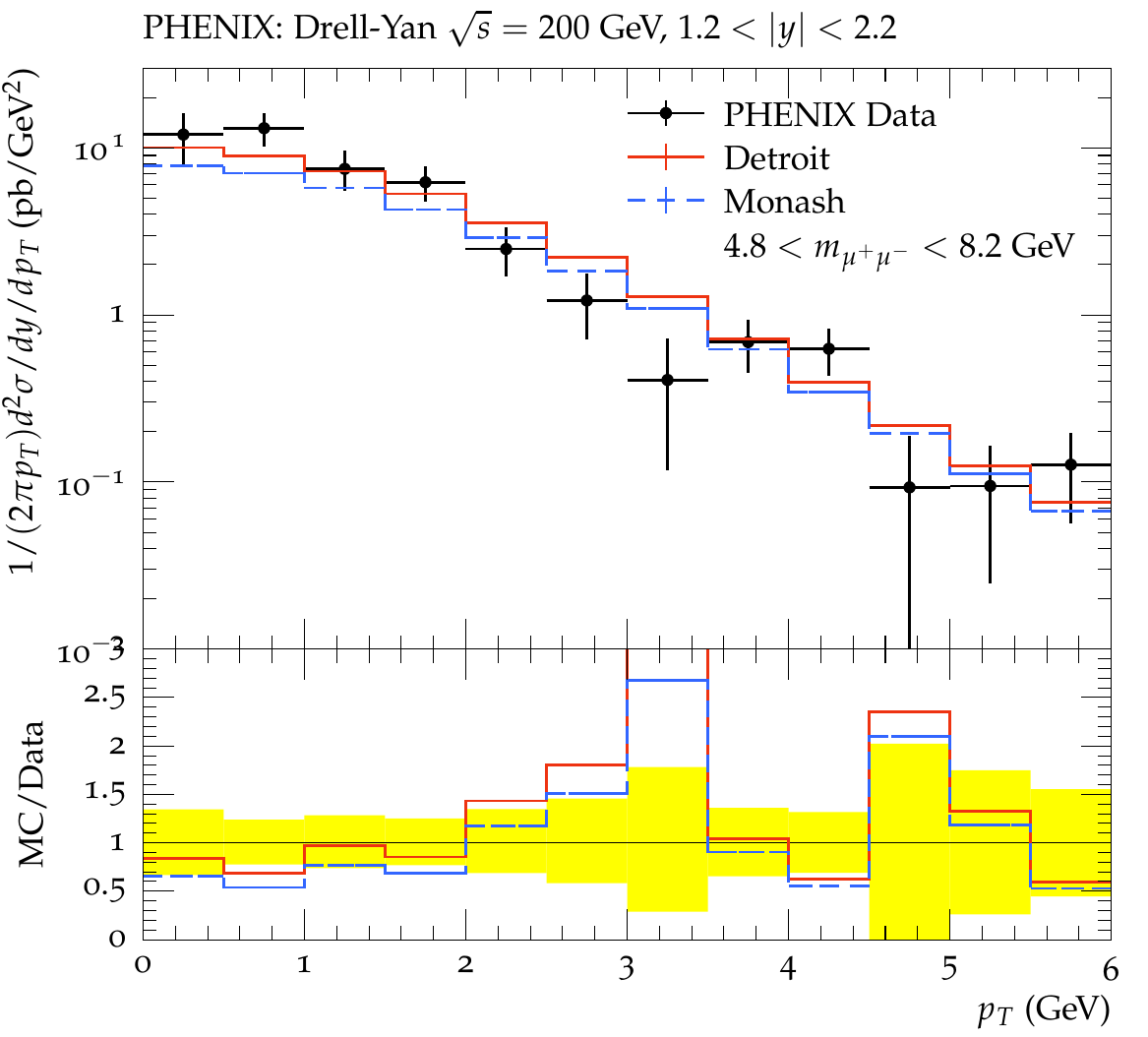}
    \includegraphics[width=0.35\textwidth]{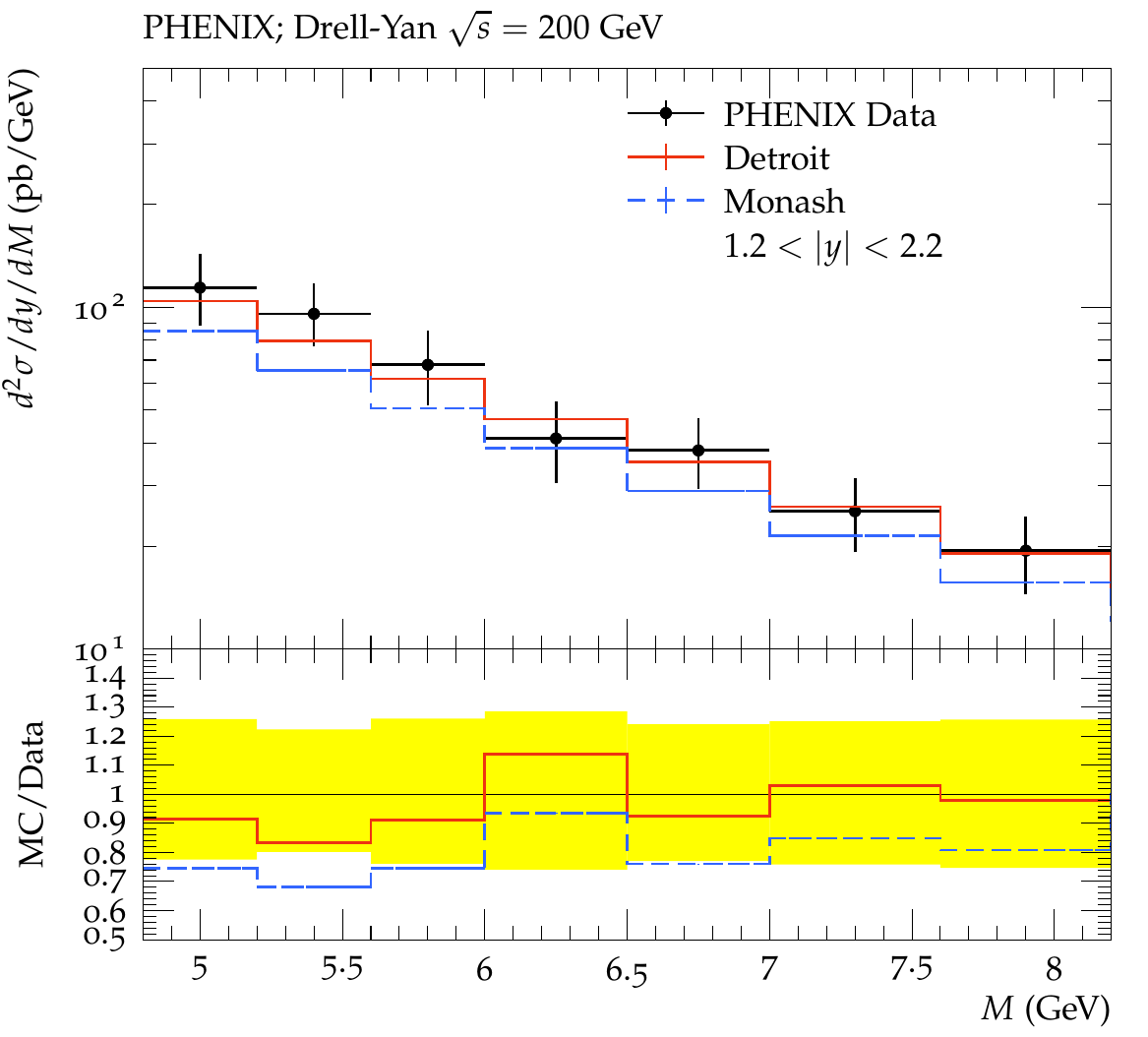}  
    \caption{Di-muon from Drell-Yan $p_{T}$ (left) and mass (right) in $p$+$p$ collisions at $\s$ = 200 GeV measured by the PHENIX experiment~\cite{PhysRevD.99.072003}. The bottom panels in each figure show the ratios of the Monte Carlo predictions with respect to the data.}
    \label{fig:dy}
\end{figure*}

\end{document}